\documentclass[12pt,a4paper,english,bibliography=totoc,index=totoc,BCOR7.5mm,titlepage,captions=tableheading]{scrartcl}
\usepackage[T1]{fontenc}
\usepackage[latin9]{inputenc}
\usepackage{color}
\usepackage{babel}
\usepackage{pifont}
\usepackage{float}
\usepackage{textcomp}
\usepackage{mathtools}
\usepackage{amsmath}
\usepackage{amssymb}
\usepackage{graphicx}
\usepackage{setspace}
\usepackage[authoryear]{natbib}
\usepackage[unicode=true,
 bookmarks=true,bookmarksnumbered=true,bookmarksopen=true,bookmarksopenlevel=1,
 breaklinks=false,pdfborder={0 0 1},backref=false,colorlinks=true]
 {hyperref}
\hypersetup{pdftitle={LyX's Math Manual},
 pdfauthor={LyX Team, Uwe Stöhr},
 pdfsubject={LyX-documentation about math},
 pdfkeywords={LyX, Mathed},
 linkcolor=black, citecolor=black, urlcolor=blue, filecolor=blue, pdfpagelayout=OneColumn, pdfnewwindow=true, pdfstartview=XYZ, plainpages=false}
\usepackage{breakurl}

\makeatletter

\providecommand{\tabularnewline}{\\}

\usepackage{ifpdf}
\ifpdf

\IfFileExists{lmodern.sty}
 {\usepackage{lmodern}}{}

\fi % end if pdflatex is used

\renewcommand{\l@subsection}{\@dottedtocline{2}{1.5em}{2.8em}}
\renewcommand{\l@subsubsection}{\@dottedtocline{3}{4.3em}{3.6em}}

\AtBeginDocument{\renewcommand{\ref}[1]{\mbox{\autoref{#1}}}}
\@ifundefined{extrasenglish}{\usepackage[english]{babel}}{}
\addto\extrasenglish{%
 \renewcommand*{\equationautorefname}[1]{}%
}

\@ifundefined{textcolor}{\usepackage{color}}{}

\let\myTOC\tableofcontents
\renewcommand{\tableofcontents}{%
 \vspace{1cm}
 \pdfbookmark[1]{\contentsname}{}
 \myTOC
 \cleardoublepage
 \pagenumbering{arabic}}

\let\myFoot\footnote
\renewcommand{\footnote}[1]{\myFoot{#1\vspace{1.5mm}}}

\setkomafont{captionlabel}{\bfseries}

\usepackage{calc}

\usepackage{multicol}

\usepackage{remreset}

\usepackage{mathrsfs}

\definecolor{darkgreen}{cmyk}{0.5, 0, 1, 0.5}

\usepackage{ifthen}

\newboolean{undertilde}
\IfFileExists{undertilde.sty}
 {\usepackage{undertilde}
  \setboolean{undertilde}{true}}
 {\setboolean{undertilde}{false}}

\newboolean{eurosym}
\IfFileExists{eurosym.sty}
 {\usepackage[gennarrow]{eurosym}
  \setboolean{eurosym}{true}}
 {\setboolean{eurosym}{false}}

\newboolean{braket}
\IfFileExists{braket.sty}
 {\usepackage{braket}
  \setboolean{braket}{true}}
 {\setboolean{braket}{false}}

\newboolean{cancel}
\IfFileExists{cancel.sty}
 {\usepackage{cancel}
  \setboolean{cancel}{true}}
 {\setboolean{cancel}{false}}

\newboolean{upgreek}
\IfFileExists{upgreek.sty}
 {\usepackage{upgreek}
  \setboolean{upgreek}{true}}
 {\setboolean{upgreek}{false}}

\AtBeginDocument{

}

\makeatother

\begin{document}

\title{Composite Inference for Gaussian Processes}

\author{Yongxiang Li\thanks{Yongxiang~Li: Department of Systems Engineering and Engineering Management,
City University of Hong Kong, email: \protect\protect\href{mailto:yongxili-c@my.cityu.edu.hk}{yongxili-c@my.cityu.edu.hk}/\protect\href{mailto:novern.li@gmail.com}{novern.li@gmail.com}. }, Qiang Zhou\thanks{Qiang ZHOU: Department of Systems and Industrial Engineering, The
University of Arizona, e-mail: \protect\protect\href{mailto:zhouq@email.arizona.edu}{zhouq@email.arizona.edu}.}, Kwok Leung~Tsui\thanks{Kwok Leung~Tsui: Department of Systems Engineering and Engineering
Management, City University of Hong Kong.}, and Javier Cabrera\thanks{Javier Cabrera: Department of Statistics, Rutgers University.} }

\maketitle
\tableofcontents{}

\noindent \begin{center}
\textbf{\Large{}Composite Inference for Gaussian Processes}{\Large{} }
\par\end{center}{\Large \par}

\noindent \begin{center}
Yongxiang Li\footnote{Department of Systems Engineering and Engineering Management, City
University of Hong Kong, email: \protect\href{mailto:yongxili-c@my.cityu.edu.hk}{yongxili-c@my.cityu.edu.hk}/\href{mailto:novern.li@gmail.com}{novern.li@gmail.com}. }, Qiang Zhou\footnote{Department of Systems and Industrial Engineering, The University of
Arizona.}, Kwok Leung~Tsui\footnote{Department of Systems Engineering and Engineering Management, City
University of Hong Kong.}, and Javier Cabrera\footnote{Department of Statistics, Rutgers University.} 
\par\end{center}

\textbf{\textit{\large{}Abstract}}: Large-scale Gaussian process models
are becoming increasingly important and widely used in many areas,
such as, computer experiments, stochastic optimization via simulation,
and machine learning using Gaussian processes. The standard methods,
such as maximum likelihood estimation (MLE) for parameter estimation
and the best linear unbiased predictor (BLUP) for prediction, are
generally the primary choices in many applications. In spite of their
merits, those methods are not feasible due to intractable computation
when the sample size is huge. A novel method for the purposes of parameter
estimation and prediction is proposed to solve the computational problems
of large-scale Gaussian process based models, by separating the original
dataset into tractable subsets. This method consistently combines
parameter estimation and prediction by making full use of the dependence
among conditional densities: a statistically efficient composite likelihood
based on joint distributions of some well selected conditional densities
is developed to estimate parameters and then ``composite inference''
is coined to make prediction for an unknown input point, based on
its distributions conditional on each block subset. The proposed method
transforms the intractable BLUP into a tractable convex optimization
problem. It is also shown that the prediction given by the proposed
method, called the best linear unbiased block predictor, has a minimum
variance for a given separation of the dataset. 

\textbf{\textit{\large{}Keywords}}: Large scale, Parallel computing,
Composite likelihood, Spatial process

\newpage

\section{Introduction}

Gaussian process models are widely used in many areas, such as, computer
experiments, stochastic optimization via simulation, and machine learning
using Gaussian processes. The explosion of interest in big data has
brought huge datasets into the spotlight, which has triggered demands
for more sophisticated statistical modeling techniques and methodologies
for large-scale Gaussian processes. Researchers always face computational
problems dealing with a huge covariance matrix when making inference
from a large-scale dataset for Gaussian process based models, using
the maximum likelihood estimation (MLE) for parameter estimation and
the best linear unbiased predictor (BLUP) for prediction. Although
the standard methods, the MLE and the BLUP, are statistically more
efficient than other alternative methods, they requires intractable
computation on a huge covariance matrix when Gaussian process models
are of large scale. Therefore, there are increasing demands in finding
approximations to these standard methods for computational convenience,
while trying not to lose too much statistical efficiency. 

Some methods have been proposed attempting to address the above computational
problem. One strategy is to simplify the covariance matrix with more
easily manipulated structures. For example, covariance tapering was
used by \citet{key-1} and \citet{key-2} to yield a sparse covariance
matrix for computational convenience; a low-rank models is used to
represent the Gaussian processes in a lower-dimensional subspace so
that calculation on a much smaller covariance matrix is only required
\citep{key-4,key-3,key-5}. In spite of its merits, using a simplified
covariance matrix for Gaussian processes may lead to unnecessary efficiency
loss (such as ignoring long-range dependence in covariance tapering),
and this strategy also has limited applications. 

To take advantage of the benefits of the MLE and the BLUP, it is urgent
to develop more efficient approximations to the full likelihood, and
thus another strategy is to use the composite likelihood \citep{key-6,key-7}
for parameter estimation and prediction. The composite likelihood
was first proposed in the literature to address computational issues
in some scenarios when the full likelihood fails in parameter estimation.
\citet{key-8} proposed the composite conditional likelihood to approximate
the full likelihood, which is further developed by \citet{key-9}
to approximate the restricted likelihood, by using blocks of observations
to improve statistical efficiency. The composite marginal likelihood
was also used to approximate the full likelihood by \citet{key-11},
\citet{key-12} and \citet{key-10}. An unnegligible factor of efficiency
loss by these composite likelihood methods is that the component likelihoods
therein may not form an optimal combination to approximate the full
likelihood. 

In this paper we adopt the composite conditional likelihood for parameter
estimation. A key difference between our composite conditional likelihood
and previous ones is that we propose a sound combination of well selected
component likelihoods to approximate the full likelihood, according
to the chain rule of conditional probability. Another critical difference
is that we consider the dependence among some component likelihoods,
which is intentionally ignored in the conventional composite likelihoods,
in order to improve the approximation to the full likelihood. Due
to these tow critical differences, the statistical efficiency of the
proposed composite likelihood is increased substantially. The proposed
method puts as much data as possible into each component conditional
likelihood while still preserves the divide and conquer aspects. It
is well known that maximum composite likelihood estimates are consistent
and asymptotically normal \citep{key-6,key-7} under some regularity
conditions, similar as those in \citet{key-14} within increasing
domain asymptotics. 

Until recently, the composite likelihood was used to mitigate computational
burden for making prediction within large-scale Gaussian process models
by \citet{key-13}, where predictions based on a block composite likelihood
was developed. However, this method fails to give optimal weights
for component likelihoods, the product of which forms the composite
likelihood used for making prediction for Gaussian processes. It seems
that the framework of the composite likelihood cannot offer any tool
to calculate these weights. Instead we transform the BLUP, which needs
the inverse of a huge covariance matrix, into a convex optimization
that only requires the inverse of small-scale covariance matrices,
by considering dependence between conditional densities in order to
calculate the optimal weights. We call this method ``composite inference''. 

This method makes full use of information by bringing together all
the block subsets into a convex optimization. Furthermore, the original
covariance matrix rather than the approximate covariance matrix is
used to make prediction so the proposed method is more accurate than
those using the approximate covariance matrix. In addition, we also
find that the method making prediction using the composite likelihood
\citep{key-13} is a sub-solution of the proposed convex optimization.
Finally, the proposed method is compatible with parallel computing
during both parameter estimation and prediction, making it even more
computationally efficient to deal with today's increasingly growing
data.

The remainder of this paper is organized as follows. Section 2 introduces
notations of Gaussian processes used throughout the paper. The composite
likelihood methods are reviewed and the proposed method is developed
in the section 3. Numerical examples are given in section 4. Finally,
section 5 concludes the paper.

\section{Gaussian Process}

Before introducing the proposed method, we first present a very brief
review of Gaussian processes \citep{key-15}. Denote the $n$ distinct
input points by $\mathbf{X}=[\mathbf{x}_{1},\mathbf{x}_{2},\cdots,\mathbf{x}_{n}]^{T}$,
$\mathbf{x}_{i}\in\mathbb{R}^{p}$, and the corresponding responses
by $\mathbf{y}=[y_{1},y_{2},\cdots,y_{n}]^{T}$. The Gaussian process
model is defined as
\begin{equation}
y(\mathbf{x})=\mathbf{f}(\mathbf{x})^{T}\boldsymbol{\beta}+z(\mathbf{x}),\label{eq:model1}
\end{equation}
where $\mathbf{f}(\mathbf{x})=\left[f_{1}(\mathbf{x}),f_{2}(\mathbf{x}),\cdots,f_{q}(\mathbf{x})\right]^{T}$
is a vector of $q$ pre-specified regression functions. $\boldsymbol{\beta}=[\beta_{1},\beta_{2},\cdots,\beta_{q}]^{T}$
are unknown coefficients to be estimated. The $z(\mathbf{x})$ is
a univariate Gaussian process with zero mean and variance $\sigma^{2}$
and its correlation function is $\mathbf{Corr}(\mathbf{x},\mathbf{x}')=K_{\boldsymbol{\phi}}(\mathbf{x},\mathbf{x}')$
with unknown correlation parameters $\boldsymbol{\phi}=\left[\phi_{1},\phi_{2},\cdots,\phi_{p}\right]^{T}$.
A common choice is the squared exponential correlation function $K_{\boldsymbol{\phi}}(\mathbf{x},\mathbf{x}')=\exp\left\{ -(\mathbf{x}-\mathbf{x}')^{T}\boldsymbol{\Xi}(\mathbf{x}-\mathbf{x}')\right\} $,
where $\boldsymbol{\Xi}=\mathbf{diag}\left(\phi_{1},\phi_{2},\cdots,\phi_{p}\right)$.
Denote the correlation matrix by $\mathbf{R}=K_{\boldsymbol{\phi}}\left(\mathbf{X},\mathbf{X}\right)=\left\{ K_{\boldsymbol{\phi}}(\mathbf{x}_{i},\mathbf{x}_{j})\right\} $,
and $\mathbf{F}=\mathbf{f}\left(\mathbf{X}\right)^{T}=[\mathbf{f}(\mathbf{x}_{1}),\mathbf{f}(\mathbf{x}_{2}),\cdots,\mathbf{f}(\mathbf{x}_{n})]^{T}$.
The log-likelihood (up to an additive constant) of the Gaussian process
model is given by 
\begin{equation}
\log\mathcal{L}_{ML}\left(\boldsymbol{\beta},\sigma,\boldsymbol{\phi};\mathbf{y}\right)=-\frac{1}{2}\left(n\log\sigma^{2}+\log\left|\mathbf{R}\right|+\frac{\left(\mathbf{y}-\mathbf{F}\boldsymbol{\beta}\right)^{T}\mathbf{R}^{-1}\left(\mathbf{y}-\mathbf{F}\boldsymbol{\beta}\right)}{\sigma^{2}}\right).\label{eq:MLE}
\end{equation}
The parameters $\boldsymbol{\beta}$, $\sigma$ and $\boldsymbol{\phi}$
can be obtained by the MLE and the BLUP at a input point $\mathbf{x}^{*}$
is 

\begin{equation}
\hat{y}_{ML}(\mathbf{x}^{*})=\mathbf{f}(\mathbf{x}^{*})^{T}\boldsymbol{\beta}+K_{\boldsymbol{\phi}}\left(\mathbf{x}^{*},\mathbf{X}\right)K_{\boldsymbol{\phi}}\left(\mathbf{X},\mathbf{X}\right)^{-1}\left(\mathbf{y}-\mathbf{F}\boldsymbol{\beta}\right),\label{eq:BLUP}
\end{equation}
where $K_{\boldsymbol{\phi}}\left(\mathbf{x}^{*},\mathbf{X}\right)=[K_{\boldsymbol{\phi}}(\mathbf{x}^{*},\mathbf{x}_{1}),K_{\boldsymbol{\phi}}(\mathbf{x}^{*},\mathbf{x}_{2}),\cdots,K_{\boldsymbol{\phi}}(\mathbf{x}^{*},\mathbf{x}_{n})]$. 

Note that both parameter estimation by the MLE and prediction by the
BLUP involve intensive computation on calculating $\mathbf{R}^{-1}$
and/or $\left|\mathbf{R}\right|$, which is of $O\left(n^{3}\right)$
complexity, making them computationally intensive. What's more, it
is impossible to load the entire covariance matrix into the memory
of normal desktop computers when $n$ is large, for example, a $100000\times100000$
matrix in MATLAB requires about 74.5GB memory. These two concerns
have triggered demands for more computationally efficient and tractble
methods to replace the MLE and the BLUP for large-scale Gaussian processes.

\section{Composite Inference for Gaussian Processes}

Maximum likelihood estimation is generally the prior choice for parameter
estimation, but repeatedly exact computation of the full likelihood
is painfully prohibitive when the sample size is very large. To address
the computational problem, the composite likelihood is adopted. The
general principle of the composite likelihood is to simplify complex
dependence relationships by computing marginal or conditional likelihoods
of a subset of the variables, and then multiplying them together to
form an estimation function.

However there are numerous combinations of component likelihoods to
form the composite likelihood, and it is unclear which combination
is better. In addition, the dependence among these component likelihoods
is intentionally ignored. Therefore these two factors make the composite
likelihood less statistically efficient than the full likelihood.
To increase its statistical efficiency, we must consider the dependence
among these component likelihoods and select a reasonable combination,
while still preserving the divide and conquer aspects of the composite
likelihood. Next we will present some building blocks that capture
the dependence among component likelihoods for Gaussian processes,
which is the key to the proposed methods for estimating unknown parameters
and making predictions.

Assume the whole dataset $\left(\mathbf{X},\mathbf{y}\right)$ is
decomposed into $k$ block subsets of roughly the same size, each
containing $n_{i}$ data points $\left(\mathbf{X}_{i},\mathbf{y}_{i}\right)$
for $i=1,\cdots,k$, where $\sum_{i=1}^{k}n_{i}=n$. The sliced Latin
hypercube design (SLHD) by \citet{key-16} can be an option to generate
these subsets. Denote $y\left(\mathbf{x}\right)$ conditional on $\left\{ y\left(\mathbf{X}_{i}\right)=\mathbf{y}_{i}\right\} $
by another random variable $\varepsilon_{i}$, i.e. $\varepsilon_{i}=y\left(\mathbf{x}\right)|y\left(\mathbf{X}_{i}\right)=\mathbf{y}_{i}$.
The question is how are $\varepsilon_{i}$ and $\varepsilon_{j}$
correlated. To answer this question, we first present a theorem (the
proof is given in the appendix) showing that a conditional random
variable can be represented by another random variable without conditioning,
followed by a Corollary applied in Gaussian processes, which shows
the correlation between $\varepsilon_{i}$ and $\varepsilon_{j}$.
\begin{description}
\item [{Theorem:}] For a random vector $\boldsymbol{\epsilon}=[\epsilon_{1},\epsilon_{2},\cdots,\epsilon_{m}]^{T}$
where each $\epsilon_{i}$ is i.i.d. following standard normal distribution,
and a $m\times n$ matrix $\mathbf{A}$ such that $\mathbf{A}^{T}\mathbf{A}$
is nonsingular, then 
\[
\mathbf{a}^{T}\boldsymbol{\epsilon}|\left\{ \mathbf{A}^{T}\boldsymbol{\epsilon}=\mathbf{z}\right\} =\mathbf{a}^{T}\left(\mathbf{I}-\mathbf{A}\left(\mathbf{A}^{T}\mathbf{A}\right)^{-1}\mathbf{A}^{T}\right)\boldsymbol{\epsilon}+\mathbf{a}^{T}\mathbf{A}\left(\mathbf{A}^{T}\mathbf{A}\right)^{-1}\mathbf{z}.
\]

\end{description}
The Theorem simplifies the conditional probability into a tractable
probability without conditioning, making other operations on the conditional
probability possible. Therefore we can have the following Corollary
(the proof is also given in the appendix) showing the joint distribution
of $\left\{ \varepsilon_{i}\right\} $. The Corollary can be used
in both parameter estimation and prediction using composite likelihood
to increase statistical efficiency, because the dependence between
conditional densities are considered, and hence motivates the proposed
method. The remarks followed give the best weights for conditional
densities such that their weighted sum reaches the minimum variance.
This weighted sum is used to approximate $y\left(\mathbf{x}\right)|y\left(\mathbf{X}\right)=\mathbf{y}$
in the proposed method for the purpose of parameter estimation and
prediction.
\begin{description}
\item [{Corollary:}] For a Gaussian process defined in \eqref{eq:model1}
with covariance function given by $\Phi\left(\mathbf{x},\mathbf{x}'\right)$
and define $\varepsilon_{i}=y\left(\mathbf{x}\right)|y\left(\mathbf{X}_{i}\right)=\mathbf{y}_{i}$,
then $\varepsilon_{1},\varepsilon_{2},\cdots,\varepsilon_{k}$ follows
a multivariate normal distribution and {\small{}
\[
\begin{cases}
\mathbb{E}\left[\varepsilon_{i}\right] & =\mathbf{f}\left(\mathbf{x}\right)^{T}\boldsymbol{\beta}+\Phi\left(\mathbf{x},\mathbf{X}_{i}\right)\Phi\left(\mathbf{X}_{i},\mathbf{X}_{i}\right)^{-1}\left(\mathbf{y}_{i}-\mathbf{f}\left(\mathbf{X}_{i}\right)^{T}\boldsymbol{\beta}\right)\\
\mathbf{Cov}\left(\varepsilon_{i},\varepsilon_{j}\right) & =\Phi\left(\mathbf{x},\mathbf{x}\right)+\Phi\left(\mathbf{x},\mathbf{X}_{i}\right)\Phi\left(\mathbf{X}_{i},\mathbf{X}_{i}\right)^{-1}\Phi\left(\mathbf{X}_{i},\mathbf{X}_{j}\right)\Phi\left(\mathbf{X}_{j},\mathbf{X}_{j}\right)^{-1}\Phi\left(\mathbf{X}_{j},\mathbf{x}\right)\\
 & -\Phi\left(\mathbf{x},\mathbf{X}_{i}\right)\Phi\left(\mathbf{X}_{i},\mathbf{X}_{i}\right)^{-1}\Phi\left(\mathbf{X}_{i},\mathbf{x}\right)-\Phi\left(\mathbf{x},\mathbf{X}_{j}\right)\Phi\left(\mathbf{X}_{j},\mathbf{X}_{j}\right)^{-1}\Phi\left(\mathbf{X}_{j},\mathbf{x}\right)
\end{cases}
\]
}{\small \par}
\item [{Remarks:}] Denote $\boldsymbol{\varepsilon}=\left[\varepsilon_{1},\varepsilon_{2},\cdots,\varepsilon_{k}\right]^{T}$
and $\boldsymbol{\Sigma}=\mathbf{Cov}\left(\boldsymbol{\varepsilon},\boldsymbol{\varepsilon}\right)$,
then $\mathbf{w}^{T}\boldsymbol{\varepsilon}$, where $\mathbf{w}^{T}\boldsymbol{i}=1$,
reaches its minimum variance, $1/\boldsymbol{i}^{T}\boldsymbol{\Sigma}^{-1}\boldsymbol{i}$,
when $\mathbf{w}^{T}=\boldsymbol{\Sigma}^{-1}\boldsymbol{i}/\boldsymbol{i}^{T}\boldsymbol{\Sigma}^{-1}\boldsymbol{i}$.
We call the distribution of $\mathbf{w}^{T}\boldsymbol{\varepsilon}$
with optimal weights the best composite conditional distribution.
\end{description}

\subsection{Parameter Estimation Using The Composite Likelihood }

The composite likelihood is a computationally efficient estimating
function which approximates the full joint likelihood function by
the product of a collection of component likelihoods. \citet{key-17}
on the analysis of spatial models is one of the first to study composite
likelihood, who worked on composite conditional likelihoods, notably
pseudo-likelihood. \citet{key-6} coined the term composite likelihood
for the product of likelihoods. More details on the composite likelihood
methods can be found in \citet{key-7}. 

We first give some notations. Denote $\varepsilon_{rs}^{i}=y\left(\mathbf{x}_{rs}\right)|y\left(\mathbf{X}_{i}\right)=\mathbf{y}_{i}$,
where $\mathbf{x}_{rs}$ denotes the $s\text{th}$ sample point in
$\mathbf{X}_{r}$ and $\boldsymbol{\varepsilon}_{rs}=\left\{ \varepsilon_{rs}^{i}\right\} _{i<r}$,
where the notation $\left\{ v_{rs}^{i}\right\} _{i<r}$ denotes a
column vector $\left[v_{rs}^{1},\cdots,v_{rs}^{r-1}\right]^{T}$.
According the Theorem, $\boldsymbol{\varepsilon}_{rs}$ is normally
distributed and {\small{}
\begin{align*}
\mathbb{E}\left[\varepsilon_{rs}^{i}\right] & =\mathbf{f}\left(\mathbf{x}_{rs}\right)^{T}\boldsymbol{\beta}+K_{\boldsymbol{\phi}}\left(\mathbf{x}_{rs},\mathbf{X}_{i}\right)K_{\boldsymbol{\phi}}\left(\mathbf{X}_{i},\mathbf{X}_{i}\right)^{-1}\left(\mathbf{y}_{i}-\mathbf{F}_{i}\boldsymbol{\beta}\right)=\mu_{rs}^{i}\\
\mathbf{Cov}(\varepsilon_{rs}^{i},\varepsilon_{rs}^{j}) & =\sigma^{2}[1+K_{\boldsymbol{\phi}}(\mathbf{x}_{rs},\mathbf{X}_{i})K_{\boldsymbol{\phi}}(\mathbf{X}_{i},\mathbf{X}_{i})^{-1}K_{\boldsymbol{\phi}}(\mathbf{X}_{i},\mathbf{X}_{j})K_{\boldsymbol{\phi}}(\mathbf{X}_{j},\mathbf{X}_{j})^{-1}K_{\boldsymbol{\phi}}(\mathbf{X}_{j},\mathbf{x}_{rs})\\
- & K_{\boldsymbol{\phi}}(\mathbf{x}_{rs},\mathbf{X}_{i})K_{\boldsymbol{\phi}}(\mathbf{X}_{i},\mathbf{X}_{i})^{-1}K_{\boldsymbol{\phi}}(\mathbf{X}_{i},\mathbf{x}_{rs})-K_{\boldsymbol{\phi}}(\mathbf{x}_{rs},\mathbf{X}_{j})K_{\boldsymbol{\phi}}(\mathbf{X}_{j},\mathbf{X}_{j})^{-1}K_{\boldsymbol{\phi}}(\mathbf{X}_{j},\mathbf{x}_{rs})]\\
 & =\sigma^{2}K_{rs}^{ij}\ ,
\end{align*}
}where $\mathbf{F}_{i}=\mathbf{f}\left(\mathbf{X}_{i}\right)^{T}$
for $i=1,2,\cdots,k$. According to the remarks of the Corollary,
the best weight is $\mathbf{w}_{rs}=\boldsymbol{i}^{T}\boldsymbol{\Sigma}_{rs}^{-1}/\boldsymbol{i}^{T}\boldsymbol{\Sigma}_{rs}^{-1}\boldsymbol{i}=\mathbf{K}_{rs}^{-1}\boldsymbol{i}/\boldsymbol{i}^{T}\mathbf{K}_{rs}^{-1}\boldsymbol{i}$,
where $\boldsymbol{\Sigma}_{rs}=\sigma^{2}\mathbf{K}_{rs}=\sigma^{2}\left\{ K_{rs}^{ij}\right\} _{i<r,j<r}$
is the covariance matrix of $\boldsymbol{\varepsilon}_{rs}$.

There are two classes of composite likelihoods commonly used in literature:
the composite conditional likelihood (CCL) and the composite marginal
likelihood (CML). In the context of composite likelihood, for example,
the composite conditional likelihood is given by 
\begin{equation}
\mathcal{L}_{CCL}\left(\boldsymbol{\beta},\sigma,\boldsymbol{\phi};\mathbf{y}\right)=\begin{cases}
\underset{r=1}{\overset{n}{\prod}}\underset{s\neq r}{\prod}P\left(y\left(\mathbf{x}_{r}\right)=y_{r}|y\left(\mathbf{x}_{s}\right)=y_{s}\right), & \text{Pairwise}\\
\underset{i=1}{\overset{k}{\prod}}\underset{j\neq i}{\prod}P\left(y\left(\mathbf{X}_{j}\right)=\mathbf{y}_{j}|y\left(\mathbf{X}_{i}\right)=\mathbf{y}_{i}\right), & \text{Block}
\end{cases}.\label{eq:CCL}
\end{equation}
Similar composite likelihood method was proposed in \citet{key-8}
and \citet{key-9}. The composite marginal likelihood was also studied
in the literature by \citet{key-11}, \citet{key-10} and \citet{key-13},
for example, given by 
\begin{equation}
\mathcal{L}_{CML}\left(\boldsymbol{\beta},\sigma,\boldsymbol{\phi};\mathbf{y}\right)=\begin{cases}
\underset{r=1}{\overset{n}{\prod}}\underset{s\neq r}{\prod}P\left(y\left(\mathbf{x}_{r}\right)=y_{r},y\left(\mathbf{x}_{s}\right)=y_{s}\right), & \text{Pairwise}\\
\underset{i=1}{\overset{k}{\prod}}P\left(y\left(\mathbf{X}_{i}\right)=\mathbf{y}_{i}\right), & \text{Block}
\end{cases}.\label{eq:CML}
\end{equation}

It can be seen that all these composite likelihoods somehow ignore
the correlation between component likelihoods and it cannot be guaranteed
that the combinations of component likelihood in theses composite
likelihoods are better than other combinations. The clue on selecting
the optimal combination is behind in the chain rule of conditional
probability. Recall that, by the chain rule, the full likelihood can
be written as{\small{}
\begin{align*}
\mathcal{L}_{ML}\left(\boldsymbol{\beta},\sigma,\boldsymbol{\phi};\mathbf{y}\right) & =P\left(y\left(\mathbf{X}_{1}\right)=\mathbf{y}_{1}\right)P\left(y\left(\mathbf{X}_{2}\right)=\mathbf{y}_{2}|y\left(\mathbf{X}_{1}\right)=\mathbf{y}_{1}\right)\\
\times\prod_{r=3}^{k} & \prod_{s=1}^{n_{r}}P\left(y\left(\mathbf{x}_{rs}\right)=y_{rs}|y\left(\mathbf{X}_{1}\right)=\mathbf{y}_{1},\cdots,y\left(\mathbf{X}_{r-1}\right)=\mathbf{y}_{r-1},y\left(\mathbf{X}_{r}^{s-1}\right)=\mathbf{y}_{r}^{s-1}\right),
\end{align*}
}where $y_{rs}$ denotes the $s\text{th}$ sample in $\mathbf{y}_{r}$,{\small{}
$\mathbf{y}_{r}^{s-1}=\left[y_{r1},\cdots,y_{rs-1}\right]^{T}$} and
{\small{}$\mathbf{X}_{r}^{s-1}=\left\{ \mathbf{x}_{r1},\cdots,\mathbf{x}_{rs-1}\right\} $}.
The best choice is to use this combination directly, but the computation
of some component likelihoods therein is still prohibitive. 

The term $p\left(y\left(\mathbf{x}_{rs}\right)|y\left(\mathbf{X}_{1}\right)=\mathbf{y}_{1},\cdots,y\left(\mathbf{X}_{r-1}\right)=\mathbf{y}_{r-1},y\left(\mathbf{X}_{r}^{s-1}\right)=\mathbf{y}_{r}^{s-1}\right)$,
called the full conditional distribution, is therefore replaced by
the best composite conditional distribution $p\left(\mathbf{w}_{rs}^{T}\boldsymbol{\varepsilon}_{rs}\right)$,
which is of much more computational convenience, by making full use
of the joint density function of $\boldsymbol{\varepsilon}_{rs}$
to catch the correlation between $\varepsilon_{rs}^{i}$ and $\varepsilon_{rs}^{j}$.
Note that we drop the conditioning on $y\left(\mathbf{X}_{r}^{s-1}\right)=\mathbf{y}_{r}^{s-1}$
for simplicity, but it can be taken into consideration if necessary. 

The component likelihood $P\left(\mathbf{w}_{rs}^{T}\boldsymbol{\varepsilon}_{rs}=y_{rs}\right)$
takes all the variables of $\mathbf{y}_{1},\cdots,\mathbf{y}_{r-1}$
into consideration in order to increase statistical efficiency, but
only a much smaller covariance matrix $\boldsymbol{\Sigma}_{rs}$
is needed to calculate the likelihood. This may be not possible in
the previously proposed composite likelihoods using either marginal
or conditional likelihoods. What's more, it doesn't suffer from difficulties
on selection of observation and/or conditional sets, which are common
problems in the previously proposed composite likelihoods. 

Therefore the proposed composite likelihood is{\small{} 
\[
\mathcal{L}_{CI}\left(\boldsymbol{\beta},\sigma,\boldsymbol{\phi};\mathbf{y}\right)=P\left(y\left(\mathbf{X}_{1}\right)=\mathbf{y}_{1}\right)P\left(y\left(\mathbf{X}_{2}\right)=\mathbf{y}_{2}|y\left(\mathbf{X}_{1}\right)=\mathbf{y}_{1}\right)\prod_{r=3}^{k}\prod_{s=1}^{n_{r}}P\left(\mathbf{w}_{rs}^{T}\boldsymbol{\varepsilon}_{rs}=y_{rs}\right).
\]
}To simplify this composite likelihood, $\boldsymbol{\varUpsilon}_{rs}=\left\{ y_{rs}-K_{\boldsymbol{\phi}}\left(\mathbf{x}_{rs},\mathbf{X}_{i}\right)K_{\boldsymbol{\phi}}\left(\mathbf{X}_{i},\mathbf{X}_{i}\right)^{-1}\mathbf{y}_{i}\right\} _{i<r}$
and $\boldsymbol{\Gamma}_{rs}=\left\{ \mathbf{f}\left(\mathbf{x}_{rs}\right)^{T}-K_{\boldsymbol{\phi}}\left(\mathbf{x}_{rs},\mathbf{X}_{i}\right)K_{\boldsymbol{\phi}}\left(\mathbf{X}_{i},\mathbf{X}_{i}\right)^{-1}\mathbf{F}_{i}\right\} _{i<r}$
are first denoted and thus $y_{rs}-\mathbb{E}\left[\mathbf{w}_{rs}^{T}\boldsymbol{\varepsilon}_{rs}\right]=\mathbf{w}_{rs}^{T}\left(\boldsymbol{\varUpsilon}_{rs}-\boldsymbol{\Gamma}_{rs}\boldsymbol{\beta}\right)\sim\mathcal{N}\left(0,\sigma^{2}/\boldsymbol{i}^{T}\mathbf{K}_{rs}^{-1}\boldsymbol{i}\right)$.
Then by definion, it is clear that $\mathbf{y}_{1}-\mathbb{E}\left[y\left(\mathbf{X}_{1}\right)\right]=\boldsymbol{\varUpsilon}_{1}-\boldsymbol{\Gamma}_{1}\boldsymbol{\beta}\sim\mathcal{N}\left(\mathbf{0},\sigma^{2}\mathbf{K}_{1}\right)$,
and $\mathbf{y}_{2}-\mathbb{E}\left[y\left(\mathbf{X}_{2}\right)|y\left(\mathbf{X}_{1}\right)=\mathbf{y}_{1}\right]=\boldsymbol{\varUpsilon}_{2}-\boldsymbol{\Gamma}_{2}\boldsymbol{\beta}\sim\mathcal{N}\left(\mathbf{0},\sigma^{2}\mathbf{K}_{2}\right)$,
where $\boldsymbol{\varUpsilon}_{1}=\mathbf{y}_{1}$, $\boldsymbol{\varUpsilon}_{2}=\mathbf{y}_{2}-K_{\boldsymbol{\phi}}\left(\mathbf{X}_{2},\mathbf{X}_{1}\right)K_{\boldsymbol{\phi}}\left(\mathbf{X}_{1},\mathbf{X}_{1}\right)^{-1}\mathbf{y}_{1}$,
$\boldsymbol{\Gamma}_{1}=\mathbf{F}_{1}$, $\boldsymbol{\Gamma}_{2}=\mathbf{F}_{2}-K_{\boldsymbol{\phi}}\left(\mathbf{X}_{2},\mathbf{X}_{1}\right)K_{\boldsymbol{\phi}}\left(\mathbf{X}_{1},\mathbf{X}_{1}\right)^{-1}\mathbf{F}_{1}$,
$\mathbf{K}_{1}=K_{\boldsymbol{\phi}}\left(\mathbf{X}_{1},\mathbf{X}_{1}\right)$
and $\mathbf{K}_{2}=K_{\boldsymbol{\phi}}\left(\mathbf{X}_{2},\mathbf{X}_{2}\right)-K_{\boldsymbol{\phi}}\left(\mathbf{X}_{2},\mathbf{X}_{1}\right)K_{\boldsymbol{\phi}}\left(\mathbf{X}_{1},\mathbf{X}_{1}\right)^{-1}K_{\boldsymbol{\phi}}\left(\mathbf{X}_{1},\mathbf{X}_{2}\right)$.
so the proposed composite log-likelihood, up to an additive constant,
becomes 

\begin{equation}
\ell_{CI}\left(\boldsymbol{\beta},\sigma,\boldsymbol{\phi};\mathbf{y}\right)=\underset{i=1}{\overset{2}{\sum}}\ell_{i}\left(\boldsymbol{\beta},\sigma,\boldsymbol{\phi};\mathbf{y}\right)+\underset{r=3}{\overset{k}{\sum}}\underset{s=1}{\overset{n_{r}}{\sum}}\ell_{rs}\left(\boldsymbol{\beta},\sigma,\boldsymbol{\phi};\mathbf{y}\right),\label{eq:CLE}
\end{equation}
where 
\[
\begin{cases}
\ell_{i}\left(\boldsymbol{\beta},\sigma,\boldsymbol{\phi};\mathbf{y}\right) & =-\frac{1}{2}\left(n_{i}\log\sigma^{2}+\log\left|\mathbf{K}_{i}\right|+\frac{\left(\boldsymbol{\varUpsilon}_{i}-\boldsymbol{\Gamma}_{i}\boldsymbol{\beta}\right)^{T}\mathbf{K}_{i}^{-1}\left(\boldsymbol{\varUpsilon}_{i}-\boldsymbol{\Gamma}_{i}\boldsymbol{\beta}\right)}{\sigma^{2}}\right)\\
\ell_{rs}\left(\boldsymbol{\beta},\sigma,\boldsymbol{\phi};\mathbf{y}\right) & =-\frac{1}{2}\left(\log\sigma^{2}-\log\left(\boldsymbol{i}^{T}\mathbf{K}_{rs}^{-1}\boldsymbol{i}\right)+\frac{\left(\boldsymbol{\varUpsilon}_{rs}-\boldsymbol{\Gamma}_{rs}\boldsymbol{\beta}\right)^{T}\mathbf{w}_{rs}\mathbf{w}_{rs}^{T}\left(\boldsymbol{\varUpsilon}_{rs}-\boldsymbol{\Gamma}_{rs}\boldsymbol{\beta}\right)}{\sigma^{2}/\left(\boldsymbol{i}^{T}\mathbf{K}_{rs}^{-1}\boldsymbol{i}\right)}\right)
\end{cases}
\]

By equating the partial derivatives of the composite log-likelihood
with regard to $\boldsymbol{\beta}$ and $\sigma^{2}$ to zero, we
can get the estimates of $\boldsymbol{\beta}$ and $\sigma^{2}$ conditional
on $\boldsymbol{\phi}$,

\[
\begin{cases}
\hat{\boldsymbol{\beta}} & =\chi_{\boldsymbol{\Gamma}\boldsymbol{\Gamma}}^{-1}\chi_{\boldsymbol{\Gamma}\boldsymbol{\varUpsilon}}\\
\hat{\sigma}^{2} & =\chi_{\boldsymbol{\varUpsilon}\boldsymbol{\varUpsilon}}+\hat{\boldsymbol{\beta}}^{T}\chi_{\boldsymbol{\Gamma}\boldsymbol{\Gamma}}\hat{\boldsymbol{\beta}}-2\hat{\boldsymbol{\beta}}^{T}\chi_{\boldsymbol{\Gamma}\boldsymbol{\varUpsilon}}
\end{cases},
\]
where 
\[
\begin{cases}
\chi_{\boldsymbol{\Gamma}\boldsymbol{\Gamma}} & =\frac{1}{n}\left(\underset{i=1}{\overset{2}{\sum}}\boldsymbol{\Gamma}_{i}^{T}\mathbf{K}_{i}^{-1}\boldsymbol{\Gamma}_{i}+\underset{r=3}{\overset{k}{\sum}}\underset{s=1}{\overset{n_{r}}{\sum}}\boldsymbol{i}^{T}\mathbf{K}_{rs}^{-1}\boldsymbol{i}\boldsymbol{\Gamma}_{rs}^{T}\mathbf{w}_{rs}\mathbf{w}_{rs}^{T}\boldsymbol{\Gamma}_{rs}\right)\\
\chi_{\boldsymbol{\Gamma}\boldsymbol{\varUpsilon}} & =\frac{1}{n}\left(\underset{i=1}{\overset{2}{\sum}}\boldsymbol{\Gamma}_{i}^{T}\mathbf{K}_{i}^{-1}\boldsymbol{\varUpsilon}_{i}+\underset{r=3}{\overset{k}{\sum}}\underset{s=1}{\overset{n_{r}}{\sum}}\boldsymbol{i}^{T}\mathbf{K}_{rs}^{-1}\boldsymbol{i}\boldsymbol{\Gamma}_{rs}^{T}\mathbf{w}_{rs}\mathbf{w}_{rs}^{T}\boldsymbol{\varUpsilon}_{rs}\right)\\
\chi_{\boldsymbol{\varUpsilon}\boldsymbol{\varUpsilon}} & =\frac{1}{n}\left(\underset{i=1}{\overset{2}{\sum}}\boldsymbol{\varUpsilon}_{i}^{T}\mathbf{K}_{i}^{-1}\boldsymbol{\varUpsilon}_{i}+\underset{r=3}{\overset{k}{\sum}}\underset{s=1}{\overset{n_{r}}{\sum}}\boldsymbol{i}^{T}\mathbf{K}_{rs}^{-1}\boldsymbol{i}\boldsymbol{\varUpsilon}_{rs}^{T}\mathbf{w}_{rs}\mathbf{w}_{rs}^{T}\boldsymbol{\varUpsilon}_{rs}\right)
\end{cases}.
\]

By plugging $\hat{\boldsymbol{\beta}}$ and $\hat{\sigma}$ into $\ell_{CI}\left(\boldsymbol{\beta},\sigma,\boldsymbol{\phi};\mathbf{y}\right)$
in \eqref{eq:CLE}, we have  {\small{}
\begin{align*}
\ell_{CI}\left(\hat{\boldsymbol{\beta}},\hat{\sigma},\boldsymbol{\phi};\mathbf{y}\right) & =-\frac{1}{2}\left(\underset{i=1}{\overset{2}{\sum}}\left(n_{i}\log\sigma^{2}+\log\left(\mathbf{K}_{i}\right)\right)+\underset{r=3}{\overset{k}{\sum}}\underset{s=1}{\overset{n_{r}}{\sum}}\left(\log\sigma^{2}-\log\boldsymbol{i}^{T}\mathbf{K}_{rs}^{-1}\boldsymbol{i}\right)+n\right)\\
 & =-\frac{1}{2}\left(n\log\hat{\sigma}^{2}+\underset{i=1}{\overset{2}{\sum}}\log\left(\mathbf{K}_{i}\right)-\underset{r=3}{\overset{k}{\sum}}\underset{s=1}{\overset{n_{r}}{\sum}}\log\boldsymbol{i}^{T}\mathbf{K}_{rs}^{-1}\boldsymbol{i}+n\right).
\end{align*}
}Therefore $\boldsymbol{\phi}$ can be finally optimized by 
\[
\hat{\boldsymbol{\phi}}=\arg\min_{\boldsymbol{\phi}}\left\{ n\log\hat{\sigma}^{2}+\underset{i=1}{\overset{2}{\sum}}\log\left(\mathbf{K}_{i}\right)-\underset{r=3}{\overset{k}{\sum}}\underset{s=1}{\overset{n_{r}}{\sum}}\log\boldsymbol{i}^{T}\mathbf{K}_{rs}^{-1}\boldsymbol{i}\right\} .
\]

\subsection{Asymptotics for The Maximum Composite Likelihood}

Denote the score function by $\nabla\ell_{n}\left(\boldsymbol{\theta}\right)=\partial\ell_{CI}\left(\boldsymbol{\beta},\sigma,\boldsymbol{\phi};\mathbf{y}\right)/\partial\boldsymbol{\theta}$,
where $\boldsymbol{\theta}=\left\{ \boldsymbol{\beta},\sigma,\boldsymbol{\phi}\right\} $
and the Hessian matrix by $\nabla^{2}\ell_{n}\left(\boldsymbol{\theta}\right)=\partial\ell_{CI}^{2}\left(\boldsymbol{\beta},\sigma,\boldsymbol{\phi};\mathbf{y}\right)/\partial\boldsymbol{\theta}\partial\boldsymbol{\theta}^{T}$.
Let $\mathbf{J}\left(\boldsymbol{\theta}\right)=\mathbf{Var}\left(\nabla\ell_{n}\left(\boldsymbol{\theta}\right)\right)$
and $\mathbf{H}\left(\boldsymbol{\theta}\right)=-\mathbb{E}\left[\nabla^{2}\ell_{n}\left(\boldsymbol{\theta}\right)\right]$
and denote the positive square root of a positive matrix $\mathbf{P}$
by $\mathbf{P}^{1/2}$, i.e. $\mathbf{P}^{\frac{1}{2}}\left(\mathbf{P}^{\frac{1}{2}}\right)^{T}=\mathbf{P}$. 

The consistency and asymptotic normality of the maximum composite
likelihood estimators are in general ensured, under the same regularity
conditions as for the usual maximum likelihood estimators \citep{key-6},
for example, the continuity, growth and convergence conditions in
\citet{key-14} and \citep{key-18}, in the context of increasing
domain asymptotics of Gaussian processes. The idea on the analysis
of the asymptotic distribution is that, under those regularity conditions,
by Taylor expansion similar to that used in the maximum likelihood,
$\hat{\boldsymbol{\theta}}_{n}=\arg\max_{\boldsymbol{\theta}}\ell_{CI}\left(\boldsymbol{\theta};\mathbf{y}\right)$
is asymptotically normally distributed: 
\[
\mathbf{J}\left(\boldsymbol{\theta}_{0}\right)^{-\frac{1}{2}}\mathbf{H}\left(\boldsymbol{\theta}_{0}\right)\left(\hat{\boldsymbol{\theta}}_{n}-\boldsymbol{\theta}_{0}\right)\overset{d}{\longrightarrow}\mathcal{N}\left(\mathbf{0},\mathbf{I}\right).
\]

\subsection{Prediction Using The Composite Likelihood}

Composite likelihood is popularly used for parameter estimation. Indeed
it can be also used in Gaussian process model to approximate the BLUP
\eqref{eq:BLUP}. In the Gaussian process model \eqref{eq:model1},
the joint distribution of $y\left(\mathbf{x}^{*}\right)$ and $y\left(\mathbf{X}\right)$
is 
\begin{equation}
\left[\begin{array}{c}
y\left(\mathbf{x}^{*}\right)\\
y\left(\mathbf{X}\right)
\end{array}\right]\sim\mathcal{N}\left(\left[\begin{array}{c}
\mathbf{f}(\mathbf{x}^{*})^{T}\boldsymbol{\beta}\\
\mathbf{f}\left(\mathbf{X}\right)^{T}\boldsymbol{\beta}
\end{array}\right],\quad\sigma^{2}\left[\begin{array}{cc}
1 & K_{\boldsymbol{\phi}}\left(\mathbf{x}^{*},\mathbf{X}\right)\\
K_{\boldsymbol{\phi}}\left(\mathbf{X},\mathbf{x}^{*}\right) & K_{\boldsymbol{\phi}}\left(\mathbf{X},\mathbf{X}\right)
\end{array}\right]\right)\ .
\end{equation}
As shown by \citet{key-19}, the maximum likelihood estimator of $y\left(\mathbf{x}^{*}\right)$
is identical to the BLUP in \eqref{eq:BLUP}. By the similar vein,
the maximum composite likelihood predictor was developed by \citep{key-13}
to approximate the maximum likelihood estimator. 

Throughout remainder of this section, we assume true values of $\boldsymbol{\beta}$,
$\sigma^{2}$ and $\boldsymbol{\phi}$ are known, but $\hat{\boldsymbol{\beta}}$,
$\hat{\sigma}^{2}$ and $\hat{\boldsymbol{\phi}}$ are used instead
in practice. The weighted composite likelihood at an unobserved location
is 
\begin{equation}
\mathcal{L}_{CL}\left(y^{*}\right)=\prod_{i=1}^{k}P\left(y\left(\mathbf{X}_{i}\right)=\mathbf{y}_{i},y\left(\mathbf{x}^{*}\right)=y^{*}\right)^{\omega_{i}}\ ,
\end{equation}
where $\omega_{i}$ is the weight of $i\text{th}$ component likelihood
and $\underset{i=1}{\overset{k}{\sum}}\omega_{i}=1$. By differentiating
the composite likelihood $\mathcal{L}_{CL}\left(y^{*}\right)$ and
equaling its first derivatives to zero, the prediction of $y\left(\mathbf{x}^{*}\right)$
is 
\begin{align}
\hat{y}_{CL}\left(\mathbf{x}^{*}\right) & =\mathbf{f}(\mathbf{x}^{*})^{T}\boldsymbol{\beta}+\sum_{i=1}^{k}W_{i}K_{\boldsymbol{\phi}}\left(\mathbf{x}^{*},\mathbf{X}_{i}\right)K_{\boldsymbol{\phi}}\left(\mathbf{X}_{i},\mathbf{X}_{i}\right)^{-1}\left(\mathbf{y}_{i}-\mathbf{f}\left(\mathbf{X}_{i}\right)^{T}\boldsymbol{\beta}\right)\ ,\label{eq:CLP}
\end{align}
where $W_{i}=\frac{\omega_{i}}{1-K_{\boldsymbol{\phi}}\left(\mathbf{x}^{*},\mathbf{X}_{i}\right)K_{\boldsymbol{\phi}}\left(\mathbf{X}_{i},\mathbf{X}_{i}\right)^{-1}K_{\boldsymbol{\phi}}\left(\mathbf{X}_{i},\mathbf{x}^{*}\right)}\bigl/\underset{i=1}{\overset{k}{\sum}}\frac{\omega_{i}}{1-K_{\boldsymbol{\phi}}\left(\mathbf{x}^{*},\mathbf{X}_{i}\right)K_{\boldsymbol{\phi}}\left(\mathbf{X}_{i},\mathbf{X}_{i}\right)^{-1}K_{\boldsymbol{\phi}}\left(\mathbf{X}_{i},\mathbf{x}^{*}\right)}$.
Similar results was proposed in \citep{key-13} where equal weights
were given, i.e. $\omega_{i}=1/k$. 

In general, the weights can be given manually according to some criteria,
but what are their optimal values? It seems that the composite likelihood
fails to solve this problem. To answer this question, we propose the
``composite inference'', which gives analytical optimal solutions.
In fact, we will see that the predictor $\hat{y}_{CL}\left(\mathbf{x}^{*}\right)$
in \eqref{eq:CLP} is a sub-solution of the proposed method.

\subsection{Composite Inference}

For making prediction from large scale Gaussian process models, composite
likelihood sounds like a good option, but it still cannot answer the
question in the last subsection. To address this problem, we propose
the composite inference here. We denote again $\varepsilon_{*}^{i}=y\left(\mathbf{x}^{*}\right)|y\left(\mathbf{X}_{i}\right)=\mathbf{y}_{i}$
and $\boldsymbol{\varepsilon}_{*}=\left[\varepsilon_{*}^{1},\varepsilon_{*}^{2},\cdots,\varepsilon_{*}^{k}\right]^{T}$,
where $\mathbf{x}^{*}$ is an unobserved location. According to the
Corollary, the expectation for $\varepsilon_{*}^{i}$ is $\mathbb{E}\left[\varepsilon_{*}^{i}\right]$
and the covariance matrix for $\boldsymbol{\varepsilon}_{*}$ is $\boldsymbol{\Sigma}_{*}=\mathbf{Cov}\left(\boldsymbol{\varepsilon}_{*},\boldsymbol{\varepsilon}_{*}\right)$. 

The linear unbiased block predictor for $y\left(\mathbf{x}^{*}\right)$
should be a linear combination of $\varepsilon_{*}^{i}$, and thus
we have

\begin{equation}
y_{CI}\left(\mathbf{x}^{*}\right)=\boldsymbol{\omega}^{T}\boldsymbol{\varepsilon}_{*}.
\end{equation}
The rest is to calculate the weight $\boldsymbol{\omega}$ such that
the prediction $y_{CI}\left(\mathbf{x}^{*}\right)$ has the minimum
variance, by the following convex optimization 
\begin{equation}
\hat{\boldsymbol{\omega}}=\arg\min_{\boldsymbol{\omega}}\quad\boldsymbol{\omega}^{T}\boldsymbol{\Sigma}_{*}\boldsymbol{\omega}\qquad\mathrm{given}\quad\boldsymbol{\omega}^{T}\mathbf{i}=1.\label{eq:OPT}
\end{equation}
As the same results in the remarks of the Corollary, the optimal solution
of the weights is 
\[
\hat{\boldsymbol{\omega}}=\frac{1}{\boldsymbol{i}^{T}\boldsymbol{\Sigma}_{*}^{-1}\boldsymbol{i}}\boldsymbol{\Sigma}_{*}^{-1}\boldsymbol{i}.
\]

In fact, by maximizing the full likelihood of $\boldsymbol{\varepsilon}_{*}$
with regard to $y_{CI}\left(\mathbf{x}^{*}\right)$, the same result
for the prediction of $y_{CI}\left(\mathbf{x}^{*}\right)$ can be
obtained, i.e., $\hat{y}_{CI}\left(\mathbf{x}^{*}\right)=\hat{\boldsymbol{\omega}}^{T}\mathbb{E}\left[\boldsymbol{\varepsilon}_{*}\right]$.
Clearly it doesn't follows into the framework of the composite likelihood.
Indeed, the composite likelihood developed by \citep{key-13} is actually
intended to approximate the full likelihood of $\boldsymbol{\varepsilon}_{*}$.
In addition, it is not quite straight forward to see that maximizing
the full likelihood of $\boldsymbol{\varepsilon}_{*}$ will reach
the minimum variance.

Note that it can not be guaranteed that $\boldsymbol{\Sigma}_{*}$
is always nonsingular, for example, when $\mathbf{x}^{*}$ is exactly
one of $\mathbf{X}$. What's more, $\boldsymbol{\Sigma}_{*}$ will
be ill conditioned if $\mathbf{x}^{*}$ is close to any one of $\mathbf{X}$.
Therefore the optimization problem \eqref{eq:OPT} is transformed
into another convex optimization 
\begin{equation}
\boldsymbol{\omega}^{T}\boldsymbol{\Sigma}_{*}\boldsymbol{\omega}=\sigma^{2}\left(\boldsymbol{\omega}^{T}\boldsymbol{\Lambda}\boldsymbol{\omega}-2\boldsymbol{\lambda}^{T}\boldsymbol{\omega}+1\right),
\end{equation}
where 
\[
\begin{cases}
\boldsymbol{\Lambda}_{i,j} & =K_{\boldsymbol{\phi}}\left(\mathbf{x}^{*},\mathbf{X}_{i}\right)K_{\boldsymbol{\phi}}\left(\mathbf{X}_{i},\mathbf{X}_{i}\right)^{-1}K_{\boldsymbol{\phi}}\left(\mathbf{X}_{i},\mathbf{X}_{j}\right)K_{\boldsymbol{\phi}}\left(\mathbf{X}_{j},\mathbf{X}_{j}\right)^{-1}K_{\boldsymbol{\phi}}\left(\mathbf{X}_{j},\mathbf{x}^{*}\right)\\
\boldsymbol{\lambda}_{i} & =K_{\boldsymbol{\phi}}\left(\mathbf{x}^{*},\mathbf{X}_{i}\right)K_{\boldsymbol{\phi}}\left(\mathbf{X}_{i},\mathbf{X}_{i}\right)^{-1}K_{\boldsymbol{\phi}}\left(\mathbf{X}_{i},\mathbf{x}^{*}\right)
\end{cases}.
\]
If the elements of $\mathbf{X}$ are distinct, it can be proved that
the covariance matrix $K\left(\mathbf{X},\mathbf{X}\right)$ is positive
definite \citep{key-15}. According to the Proposition in the appendix,
$\boldsymbol{\Lambda}$ is also positive definite and hence is nonsingular.
By using the Lagrange multiplier, one can get 
\begin{align*}
\hat{\boldsymbol{\omega}} & =\frac{1-\boldsymbol{i}^{T}\boldsymbol{\Lambda}^{-1}\boldsymbol{\lambda}}{\boldsymbol{i}^{T}\boldsymbol{\Lambda}^{-1}\boldsymbol{i}}\boldsymbol{\Lambda}^{-1}\boldsymbol{i}+\boldsymbol{\Lambda}^{-1}\boldsymbol{\lambda}.
\end{align*}

Therefore 
\begin{align}
\hat{y}_{CI}\left(\mathbf{x}^{*}\right) & =\mathbf{f}\left(\mathbf{x}^{*}\right)^{T}\boldsymbol{\beta}+\sum_{i=1}^{k}\hat{\omega}_{i}K_{\boldsymbol{\phi}}\left(\mathbf{x}^{*},\mathbf{X}_{i}\right)K_{\boldsymbol{\phi}}\left(\mathbf{X}_{i},\mathbf{X}_{i}\right)^{-1}\left(\mathbf{y}_{i}-\mathbf{f}\left(\mathbf{X}_{i}\right)^{T}\boldsymbol{\beta}\right),\label{eq:BBLUP}
\end{align}
 and
\[
\mathbf{Var}\left(\hat{y}_{CI}\left(\mathbf{x}^{*}\right)\right)=\sigma^{2}\left(1+\frac{\left(1-\boldsymbol{i}^{T}\boldsymbol{\Lambda}^{-1}\boldsymbol{\lambda}\right)^{2}}{\boldsymbol{i}^{T}\boldsymbol{\Lambda}^{-1}\boldsymbol{i}}-\boldsymbol{\lambda}^{T}\boldsymbol{\Lambda}^{-1}\boldsymbol{\lambda}\right).
\]
Clearly when $\mathbf{x}^{*}$ is exactly one of $\mathbf{X}_{i}$,
$\hat{\boldsymbol{\omega}}=\boldsymbol{\Lambda}^{-1}\boldsymbol{\lambda}=\mathbf{e}_{i}$
and hence $\mathbf{Var}\left(\hat{y}_{CI}\left(\mathbf{x}^{*}\right)\right)=0$.
It is easy to see that $\hat{y}_{CI}\left(\mathbf{x}^{*}\right)$
is unbiased and it reaches the minimum variance given the partition
$\left\{ \left(\mathbf{X}_{i},\mathbf{y}_{i}\right)\right\} $. So
we call $\hat{y}_{CI}\left(\mathbf{x}^{*}\right)$ the best linear
unbiased block predictor (BLUBP). In addition, the composite predictor
$\hat{y}_{CL}\left(\mathbf{x}^{*}\right)$ in \eqref{eq:CLP} is in
fact a sub-solution of the proposed predictor, which may not be optimal.

Moreover it is important to see that, when the density of the observations
increases to infinity, the proposed predictor $\hat{y}_{CI}\left(\mathbf{x}^{*}\right)$
will converge to the BLUP, and hence preserving the same infill asymptotic
properties as the optimal predictor. This is because that each $\mathbb{E}\left[\varepsilon_{*}^{i}\right]$
converges to the BLUP and hence its weighted average $\hat{y}_{CI}\left(\mathbf{x}^{*}\right)$,
as the density of the observations increases to infinity.

\section{Examples}

Numerical examples are given in this section to demonstrate the performance
of the proposed method for both parameter estimation and prediction,
compared with that of the methods using the maximum likelihood and
the conventional composite likelihoods. In the first part, simulation
studies are conducted. In the simulations, the design points $\mathbf{X}$
are generated by the SLHD with $k$ subgroups each with $m$ design
points and the corresponding responses are sampled from a given Gaussian
process with specified parameters. In the second part, a numerical
example on Schewfel function is given to show the performance of the
proposed method applied to large-scale applications. Note that all
the methods compared use the same setting and the same partition of
the whole dataset.

\begin{figure}[h]
\noindent \centering{}\includegraphics[scale=0.3]{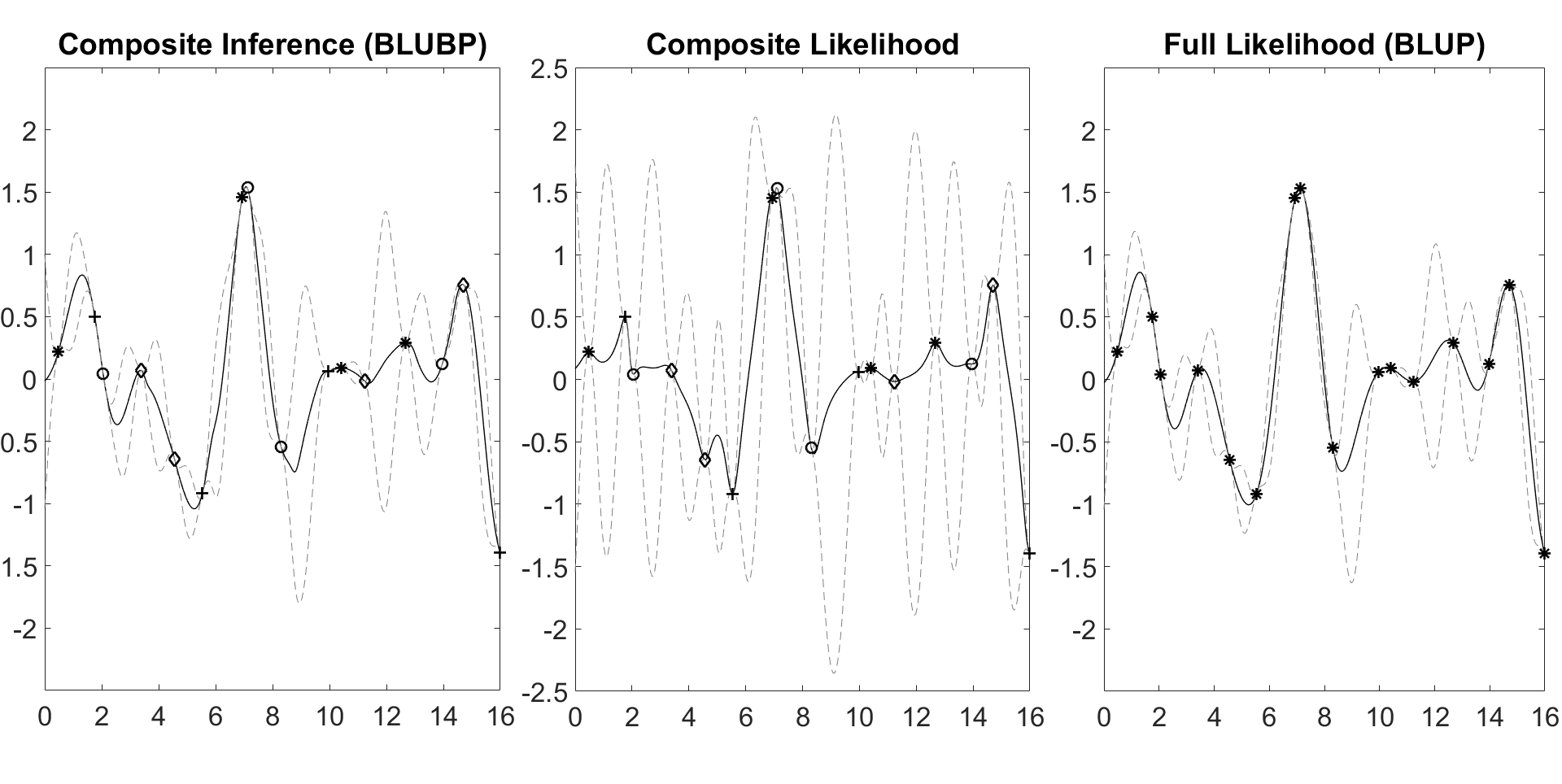}\protect\caption{\label{fig:Prediction011}Prediction from the composite inference
and the composite likelihood using 16 points with 4 subsets and that
from the full likelihood.}
\end{figure}

We first compare the composite conditional distribution, the distribution
of weighted sum of $y\left(\mathbf{x}^{*}\right)|y\left(\mathbf{X}_{i}\right)=\mathbf{y}_{i}$,
with the full conditional distribution $y\left(\mathbf{x}^{*}\right)|y\left(\mathbf{X}\right)=\mathbf{y}$.
Set $\boldsymbol{\beta}=0$, $\sigma=1$ and $\boldsymbol{\phi}=1$.
16 design points are generated by the SLHD and the Gaussian process
with those specified parameters is used to make prediction for untried
points according to the methods in comparison. Note that in this simulation
true values of parameter (rather than parameter estimates) are used
to make prediction in order to remove other distracting factors. A
typical example when $k=4$ is shown in \ref{fig:Prediction011},
where the black line is the prediction and the dashed lines are its
3-$\sigma$ confidence interval. It can be seen that the prediction
from the proposed method is very close to that from the maximum likelihood
(the BLUP), and is much better than that from the conventional composite
likelihood using equation \eqref{eq:CLP}. 

\begin{figure}[H]
\noindent \centering{}\includegraphics[scale=0.3]{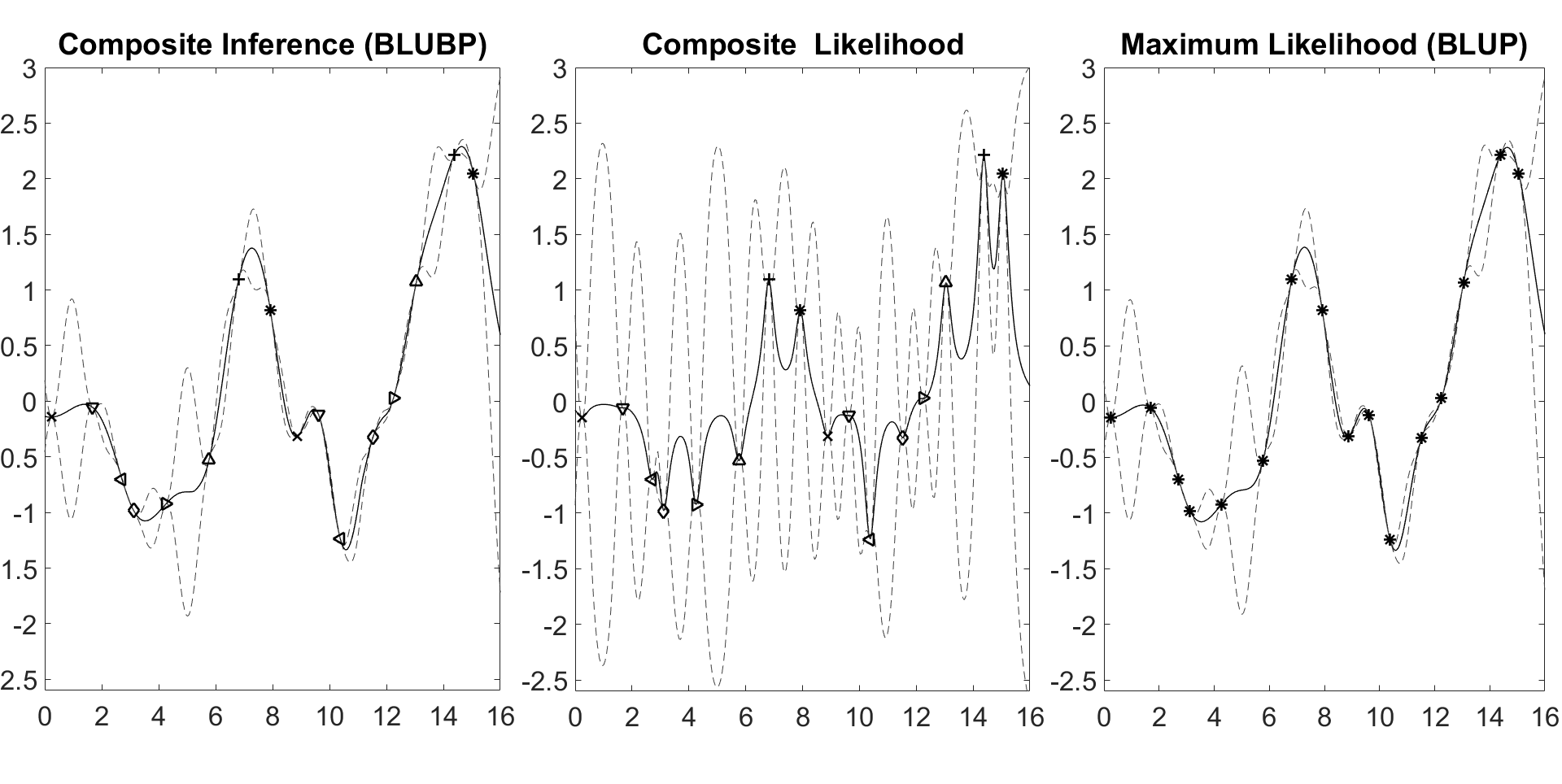}\protect\caption{\label{fig:Prediction012}Prediction from the composite inference
and the composite likelihood using 16 points with 8 subsets and that
from the full likelihood.}
\end{figure}

We increase $k$ from 4 to 8, i.e. each subgroup has 2 points, while
keeping other setting unchanged and similar results are obtained as
shown in \ref{fig:Prediction012}. It can be seen that, the proposed
method approximate extremely well to the BLUP. This suggests that
the best composite conditional distribution approximates very well
to the full conditional distribution. Simulations also shows that,
assuming the total number of design points are the same, the approximation
to the BLUP of the proposed method with larger $k$ is consistently
better than that of the proposed method with smaller $k$. In general,
the accuracy of the predictor is determined by the way of partitioning
the dataset and the information loss happens roughly in the places
of the predictor with larger uncertainty. 

Next we will compared parameter estimation using the proposed composite
likelihood (CI) with that using the the maximum likelihood (ML) and
the conventional composite likelihoods, the block version of CML and
CCL in equation \eqref{eq:CCL} and \eqref{eq:CML}. The simulation
is repeated 10000 times. In each simulation, 100 design points in
the domain $[0,100]$ are randomly generated by the SLHD with 10 subgroups
and 1000 equally space testing points are generated, and then their
responses are sampled from the given Gaussian process with $\boldsymbol{\beta}=0$,
$\sigma=1$ and $\boldsymbol{\phi}=2$. The bias and mean square error
(MSE) of parameter estimates corresponding to the 4 methods are shown
in \ref{tab:tab1d}. 
\begin{table}[H]
\noindent \begin{centering}
\begin{tabular}{|c|c|c|c|c|c|c|c|c|c|}
\hline 
 & {\scriptsize{}True} & {\scriptsize{}$Bias_{ML}$} & {\scriptsize{}$Bias_{CI}$ } & {\scriptsize{}$Bias_{CML}$} & {\scriptsize{}$Bias_{CCL}$} & {\scriptsize{}$MSE_{ML}$} & {\scriptsize{}$MSE_{CI}$} & {\scriptsize{}$MSE_{CML}$} & {\scriptsize{}$MSE_{CCL}$}\tabularnewline
\hline 
\hline 
{\scriptsize{}$\boldsymbol{\phi}$} & {\footnotesize{}2} & {\footnotesize{}0.1268} & {\footnotesize{}0.1264} & {\footnotesize{}-0.0118} & {\footnotesize{}0.1536} & {\footnotesize{}0.3577} & {\footnotesize{}0.3585} & {\footnotesize{}1.0000} & {\footnotesize{}0.4235}\tabularnewline
\hline 
{\scriptsize{}$\boldsymbol{\beta}$} & {\footnotesize{}0} & {\footnotesize{}-0.0015} & {\footnotesize{}-0.0016} & {\footnotesize{}-0.0015} & {\footnotesize{}-0.0015} & {\footnotesize{}0.0143} & {\footnotesize{}0.0143} & {\footnotesize{}0.0150} & {\footnotesize{}0.0148}\tabularnewline
\hline 
{\scriptsize{}$\sigma^{2}$} & {\footnotesize{}1} & {\footnotesize{}-0.0145} & {\footnotesize{}-0.0144} & {\footnotesize{}-0.0144} & {\footnotesize{}-0.0145} & {\footnotesize{}0.0230} & {\footnotesize{}0.0230} & {\footnotesize{}0.0243} & {\footnotesize{}0.0239}\tabularnewline
\hline 
\end{tabular}
\par\end{centering}

\protect\caption{\label{tab:tab1d}the bias and root mean square error of parameter
estimates in 1-D cases.}
\end{table}

It can be seen from \ref{tab:tab1d} that the proposed composite likelihood
gives almost the same bias and MSE of parameter estimates as the maximum
likelihood. In fact, in most simulations, it gives exactly the same
parameter estimates as the maximum likelihood. However, the MSEs of
the conventional composite likelihoods are much worse than those of
the proposed composite likelihood, because they ignores dependence
between component likelihoods and the combination of the component
likelihoods therein is not well selected. This means that the proposed
composite likelihood could be a competitive alternative to the maximum
likelihood, even for small scale Gaussian processes. For large-scale
Gaussian processes where the maximum likelihood is infeasible, we
suggest the proposed composite likelihood for parameter estimation,
especially when parallel computing services are available. 

\begin{figure}[h]
\noindent \begin{centering}
\includegraphics[scale=0.5]{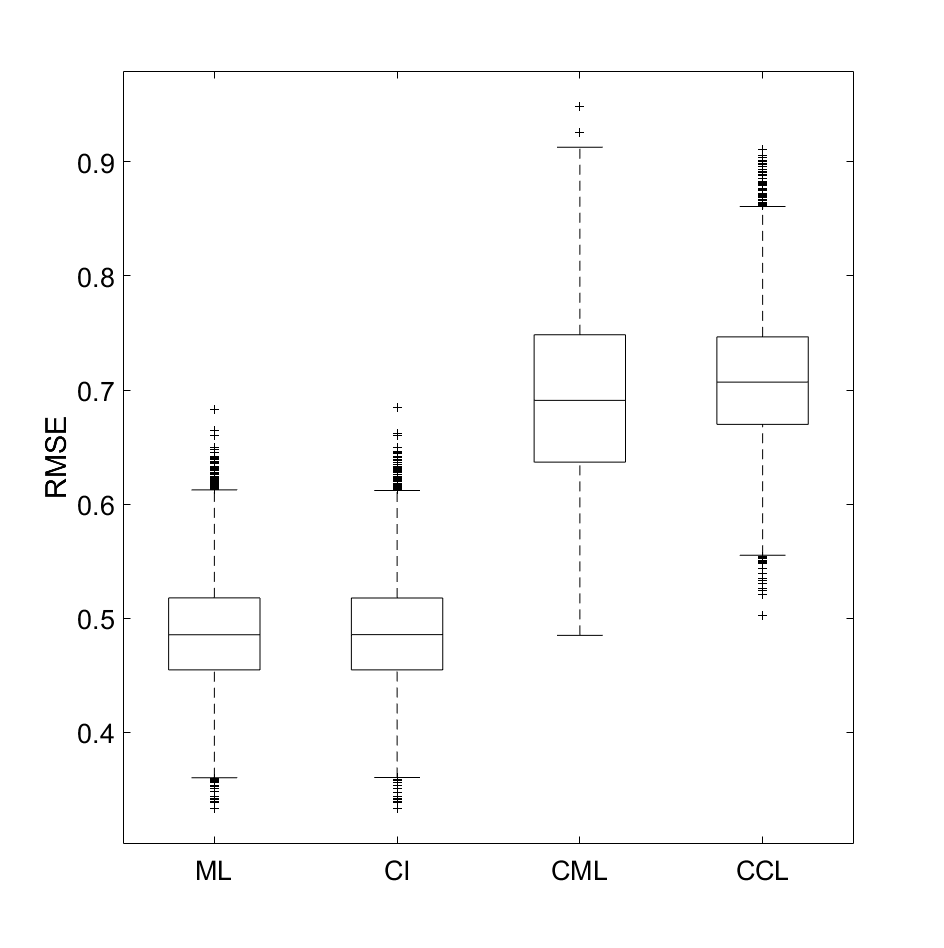}
\par\end{centering}

\protect\caption{\label{fig:Boxplots-1D}Boxplots of predictive RMSE in 1-D cases. }

\end{figure}

With optimized parameter estimates, predictions are made for the 1000
equally spaced testing points in each simulation and the RMSE of predictive
accuracy is recorded as shown in \ref{fig:Boxplots-1D}. The first
boxplot uses the standard method: the maximum likelihood for parameter
estimation and the BLUP for prediction. The proposed method for parameter
estimation and prediction is used in the second boxplot, where the
predictive accuracy is almost the same as that of the standard method.
The block version of CML and CCL is used respectively for parameter
estimation and the composite likelihood in equation \eqref{eq:CLP}
is used to make prediction in the last two boxplots, where the RMSE
of predictive accuracy is much worse than that using the proposed
method. This suggests that the proposed method performs almost the
same as the standard method, and is much better than other alternative
methods.

\begin{figure}[h]
\noindent \begin{centering}
\includegraphics[scale=0.5]{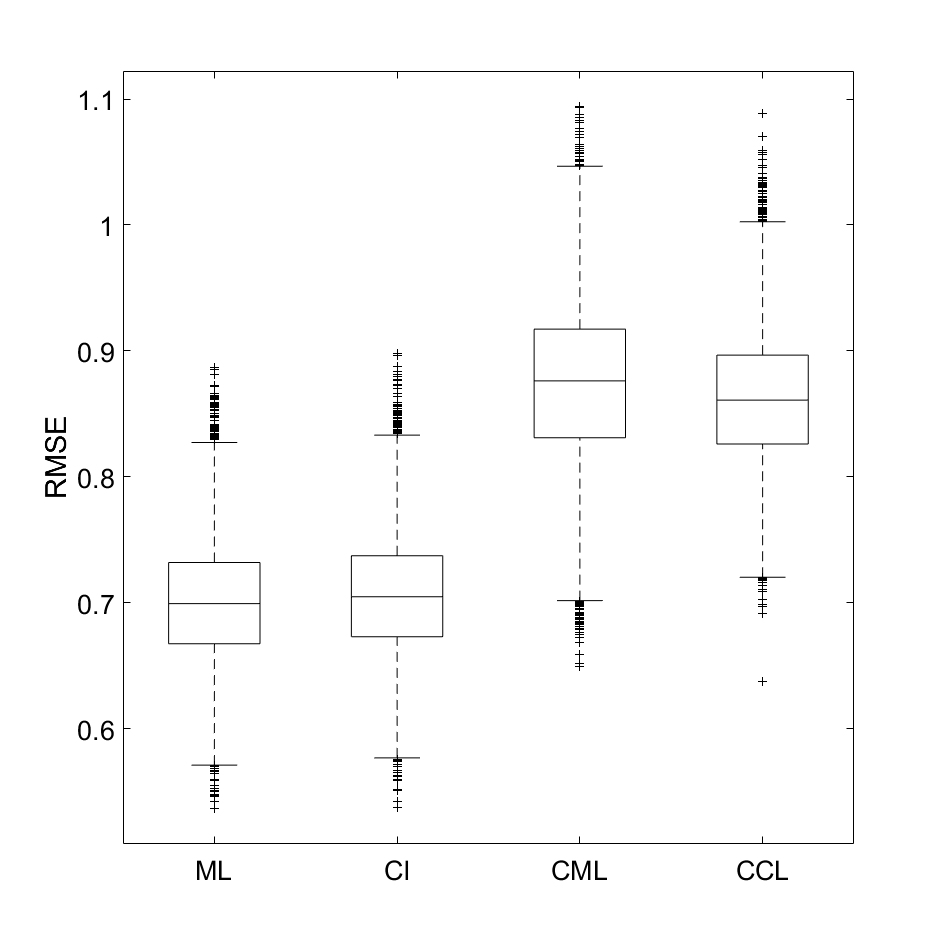}
\par\end{centering}

\protect\caption{\label{fig:Boxplots-2D}Boxplots of predictive RMSE in 2-D cases. }
\end{figure}

A simulation study in a two dimensional case is also conducted and
similar conclusions can be drawn. In each simulation, 100 design points
with 10 subgroups are randomly generated by the SLHD in the 2-D domain
$\left[0,10\right]^{2}$, with 1600 equally spaced testing points,
and their responses are generated from the Gaussian process with $\boldsymbol{\beta}=0$,
$\sigma=1$ and $\boldsymbol{\phi}=[2,2]^{T}$. The simulation is
also repeated 10000 times. The bias and MSE of parameter estimates
of each method are given in \ref{tab:tab2d}, and the predictive accuracy
is given in \ref{fig:Boxplots-2D}. 
\begin{table}[H]
\noindent \begin{centering}
\begin{tabular}{|c|c|c|c|c|c|c|c|c|c|}
\hline 
 & {\scriptsize{}True} & {\scriptsize{}$Bias_{ML}$} & {\scriptsize{}$Bias_{CI}$ } & {\scriptsize{}$Bias_{CML}$} & {\scriptsize{}$Bias_{CCL}$} & {\scriptsize{}$MSE_{ML}$} & {\scriptsize{}$MSE_{CI}$} & {\scriptsize{}$MSE_{CML}$} & {\scriptsize{}$MSE_{CCL}$}\tabularnewline
\hline 
\hline 
{\scriptsize{}$\boldsymbol{\phi}_{1}$} & {\footnotesize{}2} & {\footnotesize{}0.0632} & {\footnotesize{}0.0732} & {\footnotesize{}0.5851} & {\footnotesize{}0.1789} & {\footnotesize{}0.3519} & {\footnotesize{}0.3923} & {\footnotesize{}1.2744} & {\footnotesize{}0.6762}\tabularnewline
\hline 
{\scriptsize{}$\boldsymbol{\phi}_{2}$} & {\footnotesize{}2} & {\footnotesize{}0.0542} & {\footnotesize{}0.0648} & {\footnotesize{}0.5821} & {\footnotesize{}0.1778} & {\footnotesize{}0.3505} & {\footnotesize{}0.3953} & {\footnotesize{}1.2735} & {\footnotesize{}0.6803}\tabularnewline
\hline 
{\scriptsize{}$\boldsymbol{\beta}$} & {\footnotesize{}0} & {\footnotesize{}-0.0004} & {\footnotesize{}-0.0006} & {\footnotesize{}-0.0003} & {\footnotesize{}-0.0003} & {\footnotesize{}0.0197} & {\footnotesize{}0.0200} & {\footnotesize{}0.0230} & {\footnotesize{}0.0216}\tabularnewline
\hline 
{\scriptsize{}$\sigma^{2}$} & {\footnotesize{}1} & {\footnotesize{}-0.0101} & {\footnotesize{}-0.0117} & {\footnotesize{}-0.0201} & {\footnotesize{}-0.0179} & {\footnotesize{}0.0286} & {\footnotesize{}0.0288} & {\footnotesize{}0.0325} & {\footnotesize{}0.0310}\tabularnewline
\hline 
\end{tabular}
\par\end{centering}

\protect\caption{\label{tab:tab2d}the bias and root mean square error of parameter
estimates in 1-D cases.}
\end{table}

The parameter estimation of the proposed composite likelihood is also
extremely close to the full likelihood, though the MSE of the roughness
parameters is slightly larger. The prediction by the proposed method
is also very close to that using the standard method. On the other
hand, the parameter estimation and prediction by the conventional
composite likelihood is much worse. 

Finally, a case study on Schewfel function is given to show the performance
of the proposed method in large-scale applications compared with other
methods. The Schewfel function used here is given by 
\[
f\left(\mathbf{x}\right)=-\sum_{i=1}^{4}x_{i}\sin\sqrt{\left|1000x_{i}\right|}\ ,\text{where }-1<x_{i}<1.
\]
This function is very complex and hence 100000 data points are used
to fit a Gaussian process model. The dataset is divided into 200 subsets,
each has 500 points by the SLHD and 200000 testing points are generated
by the Latin hypercube design. The maximum likelihood fails to give
parameter estimation because it runs out of memory when calculating
the full likelihood. Therefore we compare the proposed method with
the conventional methods using composite likelihoods. The mean squared
prediction error of the proposed method is 0.1605, while the mean
squared prediction error of the methods using the conventional composite
likelihood is 0.7864 (CML) and 0.7863 (CCL) respectively. This means
that the proposed method greatly outperforms its conventional counterparts.

\section{Conclusion}

We have presented the intuitively appealing and practically useful
method on parameter estimation and prediction for Gaussian processes.
The proposed composite likelihood systematically addresses the difficulties
using composite likelihoods to approximate the full likelihood, by
considering the dependence among some well selected component likelihoods.
It is much more statistically efficient than its counterparts, and
it approximates extremely well to the full likelihood even for small-scale
Gaussian processes. The proposed composite inference gives the optimal
prediction (the BLUBP) for a given partition of the dataset, which
is also extremely close to the prediction by the BLUP. The proposed
method could be a useful and convenient alternative to the standard
method when it is not applicable. 

The proposed composite likelihood for parameter estimation is more
statistically efficient but less computationally efficient than the
conventional ones, so it can be replaced by other more computationally
efficient composite likelihoods in some scenarios where statistical
efficiency is less important than computational efficiency. On the
other hand, for making prediction, the proposed BLUBP is strongly
recommended.

The way of partitioning the whole dataset affects the prediction accuracy
of the proposed method, but it is not clear which way is the best,
and thus the best way of partitioning will be explored in the future
work. In addition, its implementation details on computational and
numerical issues, along with various examples, will be reported in
a subsequent article. Finally, the proposed method in Bayesian context
will be developed and reported elsewhere.

\section{Appendix:}
\begin{description}
\item [{Theorem:}] For a random vector $\boldsymbol{\epsilon}=[\epsilon_{1},\epsilon_{2},\cdots,\epsilon_{m}]^{T}$
where each $\epsilon_{i}$ is i.i.d. following standard normal distribution,
and a $m\times n$ matrix $\mathbf{A}$ such that $\mathbf{A}^{T}\mathbf{A}$
is nonsingular, then 
\[
\mathbf{a}^{T}\boldsymbol{\epsilon}|\left\{ \mathbf{A}^{T}\boldsymbol{\epsilon}=\mathbf{z}\right\} =\mathbf{a}^{T}\left(\mathbf{I}-\mathbf{A}\left(\mathbf{A}^{T}\mathbf{A}\right)^{-1}\mathbf{A}^{T}\right)\boldsymbol{\epsilon}+\mathbf{a}^{T}\mathbf{A}\left(\mathbf{A}^{T}\mathbf{A}\right)^{-1}\mathbf{z}\ .
\]

\item [{Proof:}] By singular value decomposition, we have $\mathbf{A}^{T}=\mathbf{U}\boldsymbol{\Lambda}\mathbf{V}$,
where $\boldsymbol{\Lambda}=\left[\begin{array}{cc}
\mathbf{D} & \mathbf{0}\end{array}\right]$ and hence $\left(\mathbf{A}^{T}\mathbf{A}\right)^{-1}=\mathbf{U}^{-T}\mathbf{D}^{-T}\mathbf{D}^{-1}\mathbf{U}^{-1}$
and $\mathbf{A}\left(\mathbf{A}^{T}\mathbf{A}\right)^{-1}\mathbf{A}^{T}=\mathbf{V}^{T}\boldsymbol{\Lambda}^{T}\mathbf{D}^{-T}\mathbf{D}^{-1}\boldsymbol{\Lambda}\mathbf{V}$.
Denote $\mathbf{e}=\mathbf{V}\boldsymbol{\epsilon}$, which is split
into two parts 
\[
\mathbf{e}=\left[\begin{array}{c}
\mathbf{e}_{1}\\
\mathbf{e}_{2}
\end{array}\right]\ ,
\]
where $\mathbf{e}_{1}$ is of size $n\times1$. Then we have 
\begin{align*}
\mathbf{a}^{T}\boldsymbol{\epsilon}|\mathbf{A}^{T}\boldsymbol{\epsilon}=\mathbf{z} & \Longleftrightarrow\mathbf{a}^{T}\mathbf{V}^{-1}\mathbf{V}\boldsymbol{\epsilon}|\mathbf{U}\boldsymbol{\Lambda}\mathbf{V}\boldsymbol{\epsilon}=\mathbf{z}\\
 & \Longleftrightarrow\mathbf{a}^{T}\mathbf{V}^{-1}\mathbf{e}|\mathbf{U}\boldsymbol{\Lambda}\mathbf{e}=\mathbf{z}\\
 & \Longleftrightarrow\mathbf{a}^{T}\mathbf{V}^{-1}\mathbf{e}|\boldsymbol{\Lambda}\mathbf{e}=\mathbf{U}^{-1}\mathbf{z}\\
 & \Longleftrightarrow\mathbf{a}^{T}\mathbf{V}^{-1}\mathbf{e}|\mathbf{D}\mathbf{e}_{1}=\mathbf{U}^{-1}\mathbf{z}\\
 & \Longleftrightarrow\mathbf{a}^{T}\mathbf{V}^{-1}\mathbf{e}|\mathbf{e}_{1}=\mathbf{D}^{-1}\mathbf{U}^{-1}\mathbf{z}\ ,
\end{align*}
and 
\begin{align*}
\mathbb{E}\left[\mathbf{a}^{T}\boldsymbol{\epsilon}|\mathbf{A}^{T}\boldsymbol{\epsilon}=\mathbf{z}\right] & =\mathbb{E}\left[\mathbf{a}^{T}\mathbf{V}^{-1}\mathbf{e}|\mathbf{e}_{1}=\mathbf{D}^{-1}\mathbf{U}^{-1}\mathbf{z}\right]\\
 & =\mathbb{E}\left[\mathbf{a}^{T}\mathbf{V}^{-1}\left[\begin{array}{c}
\mathbf{e}_{1}\\
\mathbf{e}_{2}
\end{array}\right]|\mathbf{e}_{1}=\mathbf{D}^{-1}\mathbf{U}^{-1}\mathbf{z}\right]\\
 & =\mathbf{a}^{T}\mathbf{V}^{-1}\left[\begin{array}{c}
\mathbf{D}^{-1}\mathbf{U}^{-1}\mathbf{z}\\
\mathbf{0}
\end{array}\right]\\
 & =\mathbf{a}^{T}\mathbf{V}^{-1}\left[\begin{array}{c}
\mathbf{D}\\
\mathbf{0}
\end{array}\right]\mathbf{D}^{-1}\mathbf{D}^{-1}\mathbf{U}^{-1}\mathbf{z}\\
 & =\mathbf{a}^{T}\mathbf{V}^{-1}\boldsymbol{\Lambda}^{T}\mathbf{U}^{T}\mathbf{U}^{-T}\mathbf{D}^{-1}\mathbf{D}^{-1}\mathbf{U}^{-1}\mathbf{z}\\
 & =\mathbf{a}^{T}\mathbf{V}^{T}\boldsymbol{\Lambda}^{T}\mathbf{U}^{T}\left(\mathbf{A}^{T}\mathbf{A}\right)^{-1}\mathbf{z}\\
 & =\mathbf{a}^{T}\mathbf{A}\left(\mathbf{A}^{T}\mathbf{A}\right)^{-1}\mathbf{z}\ .
\end{align*}
Therefore 
\begin{align*}
\mathbf{a}^{T}\boldsymbol{\epsilon}|\left\{ \mathbf{A}^{T}\boldsymbol{\epsilon}=\mathbf{z}\right\} -\mathbf{a}^{T}\mathbf{A}\left(\mathbf{A}^{T}\mathbf{A}\right)^{-1}\mathbf{z} & =\mathbf{a}^{T}\mathbf{V}^{-1}\left[\begin{array}{c}
\mathbf{0}\\
\mathbf{e}_{2}
\end{array}\right]\\
 & =\mathbf{a}^{T}\mathbf{V}^{-1}\left[\begin{array}{c}
\mathbf{e}_{1}\\
\mathbf{e}_{2}
\end{array}\right]-\mathbf{a}^{T}\mathbf{V}^{-1}\left[\begin{array}{c}
\mathbf{e}_{1}\\
\mathbf{0}
\end{array}\right]\\
 & =\mathbf{a}^{T}\boldsymbol{\epsilon}-\mathbf{a}^{T}\mathbf{V}^{T}\left[\begin{array}{c}
\mathbf{e}_{1}\\
\mathbf{0}
\end{array}\right]\ ,
\end{align*}
which implies 
\begin{align*}
\mathbf{a}^{T}\boldsymbol{\epsilon}|\left\{ \mathbf{A}^{T}\boldsymbol{\epsilon}=\mathbf{z}\right\}  & =\mathbf{a}^{T}\mathbf{A}\left(\mathbf{A}^{T}\mathbf{A}\right)^{-1}\mathbf{z}+\mathbf{a}^{T}\boldsymbol{\epsilon}+\mathbf{A}\left(\mathbf{A}^{T}\mathbf{A}\right)^{-1}\mathbf{A}^{T}\boldsymbol{\epsilon}\\
 & =\mathbf{a}^{T}\left(\mathbf{I}-\mathbf{A}\left(\mathbf{A}^{T}\mathbf{A}\right)^{-1}\mathbf{A}^{T}\right)\boldsymbol{\epsilon}+\mathbf{a}^{T}\mathbf{A}\left(\mathbf{A}^{T}\mathbf{A}\right)^{-1}\mathbf{z}\ ,
\end{align*}
as 
\begin{align*}
\mathbf{A}\left(\mathbf{A}^{T}\mathbf{A}\right)^{-1}\mathbf{A}^{T}\boldsymbol{\epsilon} & =\mathbf{V}^{T}\boldsymbol{\Lambda}^{T}\mathbf{D}^{-T}\mathbf{D}^{-1}\boldsymbol{\Lambda}\mathbf{V}\boldsymbol{\epsilon}\\
 & =\mathbf{V}^{T}\left[\begin{array}{c}
\mathbf{D}\\
\mathbf{0}
\end{array}\right]\mathbf{D}^{-T}\mathbf{D}^{-1}\left[\begin{array}{cc}
\mathbf{D} & \mathbf{0}\end{array}\right]\mathbf{V}\boldsymbol{\epsilon}\\
 & =\mathbf{V}^{T}\left[\begin{array}{cc}
\mathbf{I} & \mathbf{0}\\
\mathbf{0} & \mathbf{0}
\end{array}\right]\left[\begin{array}{c}
\mathbf{e}_{1}\\
\mathbf{e}_{2}
\end{array}\right]\\
 & =\mathbf{V}^{T}\left[\begin{array}{c}
\mathbf{e}_{1}\\
\mathbf{0}
\end{array}\right]\ .
\end{align*}

\item [{Corollary:}] For a Gaussian process defined in \eqref{eq:model1}
with covariance function given by $\Phi\left(\mathbf{x},\mathbf{x}'\right)=\mathbf{Cov}\left(z(\mathbf{x}),z(\mathbf{x}')\right)$
and define $\varepsilon_{i}=y\left(\mathbf{x}\right)|y\left(\mathbf{X}_{i}\right)=\mathbf{y}_{i}$,
then $\varepsilon_{1},\varepsilon_{2},\cdots,\varepsilon_{k}$ follows
a multivariate normal distribution and {\small{}
\[
\begin{cases}
\mathbb{E}\left[\varepsilon_{i}\right] & =\mathbf{f}\left(\mathbf{x}\right)^{T}\boldsymbol{\beta}+\Phi\left(\mathbf{x},\mathbf{X}_{i}\right)\Phi\left(\mathbf{X}_{i},\mathbf{X}_{i}\right)^{-1}\left(\mathbf{y}_{i}-\mathbf{f}\left(\mathbf{X}_{i}\right)^{T}\boldsymbol{\beta}\right)\\
\mathbf{Cov}\left(\varepsilon_{i},\varepsilon_{j}\right) & =\Phi\left(\mathbf{x},\mathbf{x}\right)+\Phi\left(\mathbf{x},\mathbf{X}_{i}\right)\Phi\left(\mathbf{X}_{i},\mathbf{X}_{i}\right)^{-1}\Phi\left(\mathbf{X}_{i},\mathbf{X}_{j}\right)\Phi\left(\mathbf{X}_{j},\mathbf{X}_{j}\right)^{-1}\Phi\left(\mathbf{X}_{j},\mathbf{x}\right)\\
 & -\Phi\left(\mathbf{x},\mathbf{X}_{i}\right)\Phi\left(\mathbf{X}_{i},\mathbf{X}_{i}\right)^{-1}\Phi\left(\mathbf{X}_{i},\mathbf{x}\right)-\Phi\left(\mathbf{x},\mathbf{X}_{j}\right)\Phi\left(\mathbf{X}_{j},\mathbf{X}_{j}\right)^{-1}\Phi\left(\mathbf{X}_{j},\mathbf{x}\right)
\end{cases}
\]
}{\small \par}
\item [{Proof:}]~
\end{description}
Let $\mathbf{y}_{i}=\mathbf{f}\left(\mathbf{X}_{i}\right)^{T}\boldsymbol{\beta}+\mathbf{z}_{i}$,
$\mathbf{y}_{j}=\mathbf{f}\left(\mathbf{X}_{i}\right)^{T}\boldsymbol{\beta}+\mathbf{z}_{j}$.
Since {\small{}
\[
\left[\begin{array}{c}
y\left(\mathbf{x}\right)-\mathbf{f}\left(\mathbf{x}\right)^{T}\boldsymbol{\beta}\\
y\left(\mathbf{X}_{i}\right)-\mathbf{f}\left(\mathbf{X}_{i}\right)^{T}\boldsymbol{\beta}\\
y\left(\mathbf{X}_{j}\right)-\mathbf{f}\left(\mathbf{X}_{j}\right)^{T}\boldsymbol{\beta}
\end{array}\right]=\left[\begin{array}{c}
z\left(\mathbf{x}\right)\\
z\left(\mathbf{X}_{i}\right)\\
z\left(\mathbf{X}_{j}\right)
\end{array}\right]\sim\mathcal{N}\left(\left[\begin{array}{c}
\mathbf{0}\\
\mathbf{0}\\
\mathbf{0}
\end{array}\right],\left[\begin{array}{ccc}
\Phi\left(\mathbf{x},\mathbf{x}\right) & \Phi\left(\mathbf{x},\mathbf{X}_{i}\right) & \Phi\left(\mathbf{x},\mathbf{X}_{j}\right)\\
\Phi\left(\mathbf{X}_{i},\mathbf{x}\right) & \Phi\left(\mathbf{X}_{i},\mathbf{X}_{i}\right) & \Phi\left(\mathbf{X}_{i},\mathbf{X}_{j}\right)\\
\Phi\left(\mathbf{X}_{j},\mathbf{x}\right) & \Phi\left(\mathbf{X}_{j},\mathbf{X}_{i}\right) & \Phi\left(\mathbf{X}_{j},\mathbf{X}_{j}\right)
\end{array}\right]\right)
\]
}thus there exist $\boldsymbol{\epsilon}\sim\mathcal{N}\left(\mathbf{0},\mathbf{I}\right)$
and $\mathbf{A}$, $\mathbf{A}_{j}$, $\mathbf{A}_{j}$ such that
\[
\left[\begin{array}{c}
z\left(\mathbf{x}\right)\\
z\left(\mathbf{X}_{i}\right)\\
z\left(\mathbf{X}_{j}\right)
\end{array}\right]=\left[\begin{array}{c}
\mathbf{A}^{T}\boldsymbol{\epsilon}\\
\mathbf{A}_{j}^{T}\boldsymbol{\epsilon}\\
\mathbf{A}_{j}^{T}\boldsymbol{\epsilon}
\end{array}\right]\text{, and }
\]
\[
\left[\begin{array}{ccc}
\mathbf{A}^{T}\mathbf{A} & \mathbf{A}^{T}\mathbf{A}_{i} & \mathbf{A}^{T}\mathbf{A}_{j}\\
\mathbf{A}_{i}^{T}\mathbf{A} & \mathbf{A}_{i}^{T}\mathbf{A}_{i} & \mathbf{A}_{i}^{T}\mathbf{A}_{j}\\
\mathbf{A}_{j}^{T}\mathbf{A} & \mathbf{A}_{j}^{T}\mathbf{A}_{i} & \mathbf{A}_{j}^{T}\mathbf{A}_{j}
\end{array}\right]=\left[\begin{array}{ccc}
\Phi\left(\mathbf{x},\mathbf{x}\right) & \Phi\left(\mathbf{x},\mathbf{X}_{i}\right) & \Phi\left(\mathbf{x},\mathbf{X}_{j}\right)\\
\Phi\left(\mathbf{X}_{i},\mathbf{x}\right) & \Phi\left(\mathbf{X}_{i},\mathbf{X}_{i}\right) & \Phi\left(\mathbf{X}_{i},\mathbf{X}_{j}\right)\\
\Phi\left(\mathbf{X}_{j},\mathbf{x}\right) & \Phi\left(\mathbf{X}_{j},\mathbf{X}_{i}\right) & \Phi\left(\mathbf{X}_{j},\mathbf{X}_{j}\right)
\end{array}\right]\ .
\]
By the Theorem we have 
\[
\begin{cases}
z\left(\mathbf{x}\right)|\left\{ z\left(\mathbf{X}_{i}\right)=\mathbf{z}_{i}\right\}  & =\mathbf{A}^{T}\left(\mathbf{I}-\mathbf{A}_{i}\left(\mathbf{A}_{i}^{T}\mathbf{A}_{i}\right)^{-1}\mathbf{A}_{i}^{T}\right)\boldsymbol{\epsilon}+\mathbf{A}^{T}\mathbf{A}_{i}\left(\mathbf{A}_{i}^{T}\mathbf{A}_{i}\right)^{-1}\mathbf{z}_{i}\\
z\left(\mathbf{x}\right)|\left\{ z\left(\mathbf{X}_{j}\right)=\mathbf{z}_{j}\right\}  & =\mathbf{A}^{T}\left(\mathbf{I}-\mathbf{A}_{j}\left(\mathbf{A}_{j}^{T}\mathbf{A}_{j}\right)^{-1}\mathbf{A}_{j}^{T}\right)\boldsymbol{\epsilon}+\mathbf{A}^{T}\mathbf{A}_{j}\left(\mathbf{A}_{j}^{T}\mathbf{A}_{j}\right)^{-1}\mathbf{z}_{j}
\end{cases}\ .
\]
Note that 
\[
\begin{cases}
\varepsilon_{i}=y\left(\mathbf{x}\right)|\left\{ y\left(\mathbf{X}_{i}\right)=\mathbf{y}_{i}\right\}  & =\mathbf{f}\left(\mathbf{x}\right)^{T}\boldsymbol{\beta}+z\left(\mathbf{x}\right)|\left\{ z\left(\mathbf{X}_{i}\right)=\mathbf{z}_{i}\right\} \\
\varepsilon_{j}=y\left(\mathbf{x}\right)|\left\{ y\left(\mathbf{X}_{j}\right)=\mathbf{y}_{j}\right\}  & =\mathbf{f}\left(\mathbf{x}\right)^{T}\boldsymbol{\beta}+z\left(\mathbf{x}\right)|\left\{ z\left(\mathbf{X}_{j}\right)=\mathbf{z}_{j}\right\} 
\end{cases}\ ,
\]
and thus $\varepsilon_{1},\varepsilon_{2},\cdots,\varepsilon_{k}$
follows a multivariate normal distribution and 
\[
\begin{cases}
\mathbb{E}\left[\varepsilon_{i}\right] & =\mathbf{f}\left(\mathbf{x}\right)^{T}\boldsymbol{\beta}+\mathbf{A}^{T}\mathbf{A}_{i}\left(\mathbf{A}_{i}^{T}\mathbf{A}_{i}\right)^{-1}\left(\mathbf{y}_{i}-\mathbf{f}\left(\mathbf{X}_{i}\right)^{T}\boldsymbol{\beta}\right)\\
 & =\mathbf{f}\left(\mathbf{x}\right)^{T}\boldsymbol{\beta}+\Phi\left(\mathbf{x},\mathbf{X}_{i}\right)\Phi\left(\mathbf{X}_{i},\mathbf{X}_{i}\right)^{-1}\left(\mathbf{y}_{i}-\mathbf{f}\left(\mathbf{X}_{i}\right)^{T}\boldsymbol{\beta}\right)\\
\mathbf{Cov}\left(\varepsilon_{i},\varepsilon_{j}\right) & =\mathbf{A}^{T}\left(\mathbf{I}-\mathbf{A}_{i}\left(\mathbf{A}_{i}^{T}\mathbf{A}_{i}\right)^{-1}\mathbf{A}_{i}^{T}\right)\left(\mathbf{I}-\mathbf{A}_{j}\left(\mathbf{A}_{j}^{T}\mathbf{A}_{j}\right)^{-1}\mathbf{A}_{j}^{T}\right)\mathbf{A}\\
 & =\Phi\left(\mathbf{x},\mathbf{x}\right)+\Phi\left(\mathbf{x},\mathbf{X}_{i}\right)\Phi\left(\mathbf{X}_{i},\mathbf{X}_{i}\right)^{-1}\Phi\left(\mathbf{X}_{i},\mathbf{X}_{j}\right)\Phi\left(\mathbf{X}_{j},\mathbf{X}_{j}\right)^{-1}\Phi\left(\mathbf{X}_{j},\mathbf{x}\right)\\
 & -\Phi\left(\mathbf{x},\mathbf{X}_{i}\right)\Phi\left(\mathbf{X}_{i},\mathbf{X}_{i}\right)^{-1}\Phi\left(\mathbf{X}_{i},\mathbf{x}\right)-\Phi\left(\mathbf{x},\mathbf{X}_{j}\right)\Phi\left(\mathbf{X}_{j},\mathbf{X}_{j}\right)^{-1}\Phi\left(\mathbf{X}_{j},\mathbf{x}\right)
\end{cases}\ .
\]

\begin{description}
\item [{Proposition:}] Denote $\mathbf{X}=[\mathbf{x}_{1},\mathbf{x}_{2},\cdots,\mathbf{x}_{n}]^{T}$,
which is divided into $k$ subsets: $\mathbf{X}_{1}\cdots,\mathbf{X}_{k}$.
If $K(\mathbf{x},\mathbf{x}')$ be the correlation function of two
points $\mathbf{x}$ and $\mathbf{x}'$, such that $K\left(\mathbf{X},\mathbf{X}\right)$
is positive definite, then $\mathbf{U}$ is positive definite, where
\[
\mathbf{U}_{i,j}=K\left(\mathbf{x}^{*},\mathbf{X}_{i}\right)K\left(\mathbf{X}_{i},\mathbf{X}_{i}\right)^{-1}K\left(\mathbf{X}_{i},\mathbf{X}_{j}\right)K\left(\mathbf{X}_{j},\mathbf{X}_{j}\right)^{-1}K\left(\mathbf{X}_{j},\mathbf{x}^{*}\right)\ .
\]

\item [{Proof:}] We can always find $\mathbf{A}=\left[\mathbf{A}_{1},\cdots,\mathbf{A}_{k}\right]$,
such that $\mathbf{A}^{T}\mathbf{A}=K\left(\mathbf{X},\mathbf{X}\right)$,
which means that $\mathbf{A}_{i}^{T}\mathbf{A}_{j}=K\left(\mathbf{X}_{i},\mathbf{X}_{j}\right)$.
As $K\left(\mathbf{X},\mathbf{X}\right)$ is strictly positive definite,
then $\mathrm{rank}\left(\mathbf{A}\right)=n$. Thus $\mathbf{U}=\mathbf{B}^{T}\mathbf{B},$
where $\mathbf{B}=[\mathbf{A}_{1}\mathbf{r}_{1},\ \cdots,\ \mathbf{A}_{k}\mathbf{r}_{k}],$
where $\mathbf{r}_{i}=K\left(\mathbf{X}_{i},\mathbf{X}_{i}\right)^{-1}K\left(\mathbf{X}_{i},\mathbf{x}^{*}\right)$.
If $\mathrm{rank}\left(\mathbf{B}\right)<k$, i.e. the $k$ column
vectors of $\mathbf{B}$ are linear-dependent, then there exists $\omega_{1},\cdots,\omega_{k}$,
not all zero, such that 
\[
\sum_{i=1}^{k}\mathbf{A}_{i}\mathbf{r}_{i}\omega_{i}=\mathbf{0}\ ,
\]
which implies $\mathbf{A}\boldsymbol{\omega}=\mathbf{0}$, where 
\[
\boldsymbol{\omega}=\left[\begin{array}{c}
\mathbf{r}_{1}\omega_{1}\\
\vdots\\
\mathbf{r}_{i}\omega_{k}
\end{array}\right]\ ,
\]
 not all zero. This contradicts with $\mathrm{rank}\left(\mathbf{A}\right)=n$.
Therefore $\mathbf{U}$ is positive definite.
\end{description}
\newpage{}

\end{document}